\documentclass[prd,twocolumn,superscriptaddress]{revtex4-1}

\usepackage{amsmath}
\usepackage{epsfig}
\usepackage{graphicx}
\usepackage{color}
\usepackage[normalem]{ulem}
\usepackage{amsfonts}
\usepackage[breaklinks, plainpages=false, colorlinks=true, anchorcolor=cyan, linkcolor=red, citecolor=cyan, urlcolor=magenta, bookmarks=false]{hyperref}

\usepackage[caption=false]{subfig}

\newcommand{\gparams}{\mathbf\theta} 
\newcommand{\params}{\mathbf\lambda} 
\newcommand{\Params}{\mathbf\Lambda}
\newcommand{\amp}{\mathcal{A}}
\newcommand{\f}{f_0}
\newcommand{\phase}{\varphi_0}
\newcommand{\fdot}{\dot{f}}
\newcommand{\SNR}{\rm{SNR}}
\newcommand{\BayesLine}{\texttt{BayesLine}}
\newcommand{\BayesWave}{\texttt{BayesWave}}

\newcommand{\data}{{\bf d}}

\newcommand{\Nparams}{N_{\rm P}}
\newcommand{\Nsources}{N_{\rm GW}}
\newcommand{\template}{{\boldsymbol h}}
\newcommand{\templatesum}{{\mathbf h}}
\newcommand{\innerproduct}[2]{\langle #1 | #2 \rangle}

\begin{document}
\renewcommand{\thefigure}{\arabic{figure}}
\setcounter{figure}{0}

 \def\I{{\rm i}}
 \def\E{{\rm e}}
 \def\D{{\rm d}}

\bibliographystyle{apsrev}

\title{Global Analysis of the Gravitational Wave Signal from Galactic Binaries}

\author{Tyson B. Littenberg}
\affiliation{NASA Marshall Space Flight Center, Huntsville, Alabama 35811, USA}

\author{Neil J. Cornish}
\affiliation{eXtreme Gravity Institute, Department of Physics, Montana State University, Bozeman, Montana 59717, USA}
\affiliation{Artemis, Universit\'{e} C\^{o}te dÃAzur, Observatoire C\^{o}te dÃAzur, CNRS, CS 34229, F-06304 Nice Cedex 4, France}

\author{Kristen Lackeos}
\affiliation{NASA Postdoctoral Program Fellow, NASA Marshall Space Flight Center, Huntsville, Alabama 35812, USA}

\author{Travis Robson}
\affiliation{eXtreme Gravity Institute, Department of Physics, Montana State University, Bozeman, Montana 59717, USA}

\begin{abstract} 
Galactic ultra compact binaries are expected to be the dominant source of gravitational waves in the milli-Hertz frequency band. Of the tens of millions of galactic binaries with periods shorter than an hour, it is estimated that a few tens of thousand will be resolved by the future Laser Interferometer Space Antenna (LISA). The unresolved remainder will be the main source of ``noise'' between 1-3 milli-Hertz. Typical galactic binaries are millions of years from merger, and consequently their signals will persist for the the duration of the LISA mission. Extracting tens of thousands of overlapping galactic signals and characterizing the unresolved component is a central challenge in LISA data analysis, and a key contribution to arriving at a global solution that simultaneously fits for all signals in the band. Here we present an end-to-end analysis pipeline for galactic binaries that uses trans-dimensional Bayesian inference to develop a time-evolving catalog of sources as data arrive from the LISA constellation.

\end{abstract}

\maketitle

\section{Introduction}
The most prolific source of gravitational waves (GWs) in the mHz band are galactic ultra compact binaries (UCBs), primarily comprised of two white dwarf stars. 
Ref.~\cite{Korol:2017qcx} describes a contemporary prediction for the population of UCBs detectable by the Laser Interferometer Space Antenna (LISA)~\cite{LISA}.
GWs from UCBs are continuous sources for LISA, several thousands of which will be individually resolvable. The remaining binaries blend together to form a confusion-limited foreground that is expected to be the dominant ``noise'' contribution to the LISA data stream at frequencies below ${\sim}3$ mHz, the extent of which depending on the population of binaries and the observing time of LISA~\cite{Cornish:2017vip}.

Of the thousands of resolvable binaries, the best-measured systems will serve as laboratories for studying the dynamical evolution of the binaries. 
Encoded within the orbital dynamics are relativistic effects, the internal structure of WD stars, and effects of mass transfer~\cite{Taam1980,Savonije1986,Willems2008,Nelemans2010,Littenberg_2019, Piro_2019}.
The observable population of UCBs will depend on astrophysical processes undergone by binary stars that are currently not well understood, including the formation of the compact objects themselves, binary evolution, and the end result for such binaries~\citep{Webbink1984}.
UCBs are detectable anywhere in the galaxy because the GW signals are unobscured by intervening material in the Galactic plane, providing an unbiased sample to infer large scale structure of the Milky Way~\cite{Adams:2012qw,Korol2018}.
While LISA will dramatically increase our understanding of UCBs in the galaxy, there is an ever-increasing number of systems discovered by electromagnetic (EM) observations that will be easily detectable by LISA~\cite{Kupfer_2018, Burdge_2019,Burdge_2019b, Brown_2020}. Thus UCBs are guaranteed multimessneger sources and the joint EM+GW observations provide physical constraints on masses, radii, and orbital dynamics far beyond what independent EM or GW observations can achieve alone~\cite{Shah2014a, Littenberg_2019b}.

The optimal detection, characterization, and removal of UCBs from the data stream has been long recognized as a fundamentally important and challenging aspect of the broader LISA analysis. Over-fitting the galaxy will result in a large contamination fraction in the catalog of detected sources, while under-fitting the UCB population will degrade the analyses of extragalactic sources in the data due to the excess residual. 

In this paper we describe a modern implementation of a UCB analysis pipeline which is a direct descendent of the trailblazing algorithms designed in response to the original Mock LISA Data Challenges (MLDCs)~\cite{Babak_2008,Babak_2010}, and similar methods developed for astrophysical transients and non-Gaussian detector noise currently in use for ground-based GW observations~\cite{Cornish:2014kda,Littenberg:2015}.

\begin{figure*}[htp]
\includegraphics[width=0.45\textwidth]{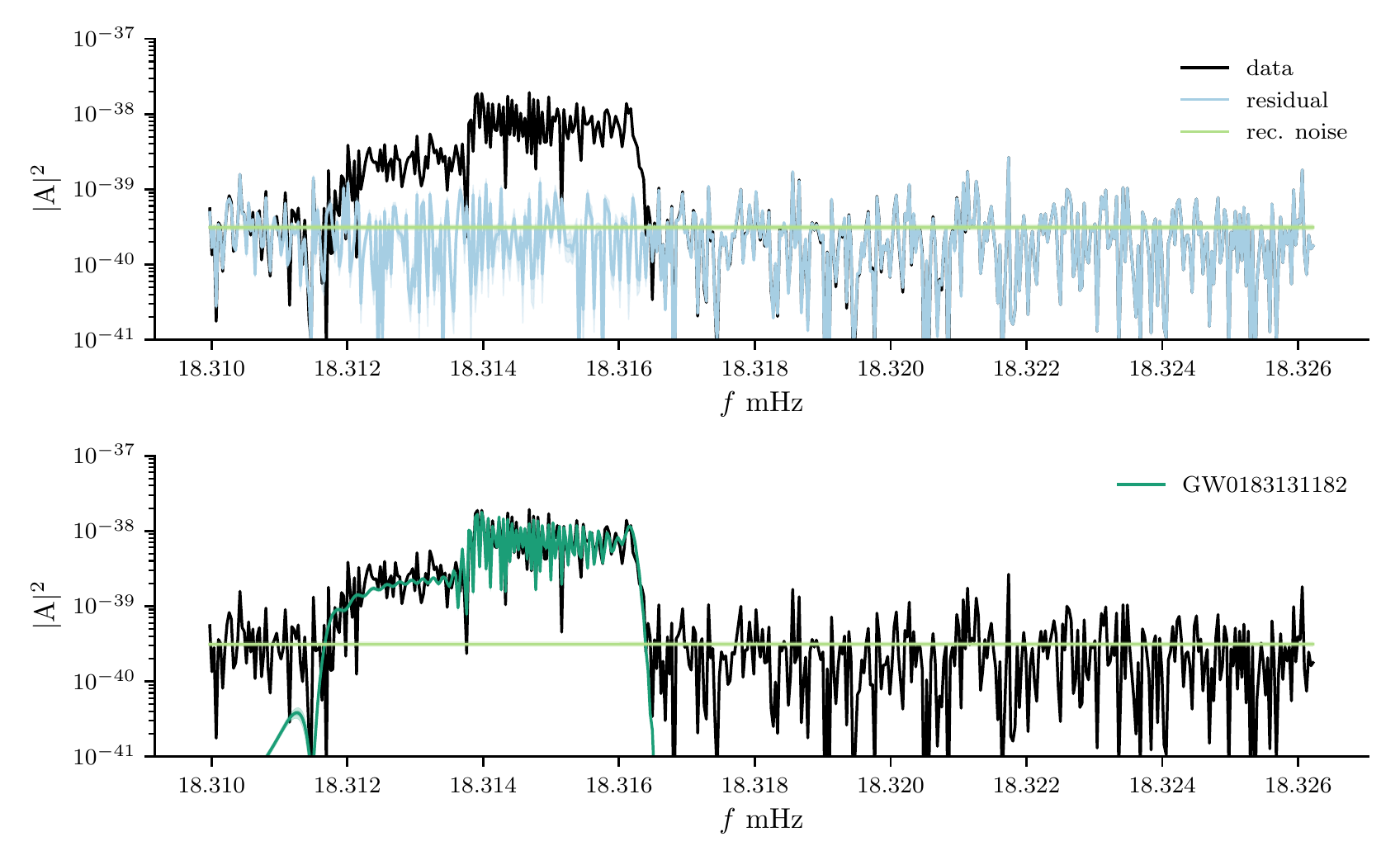} 
\includegraphics[width=0.45\textwidth]{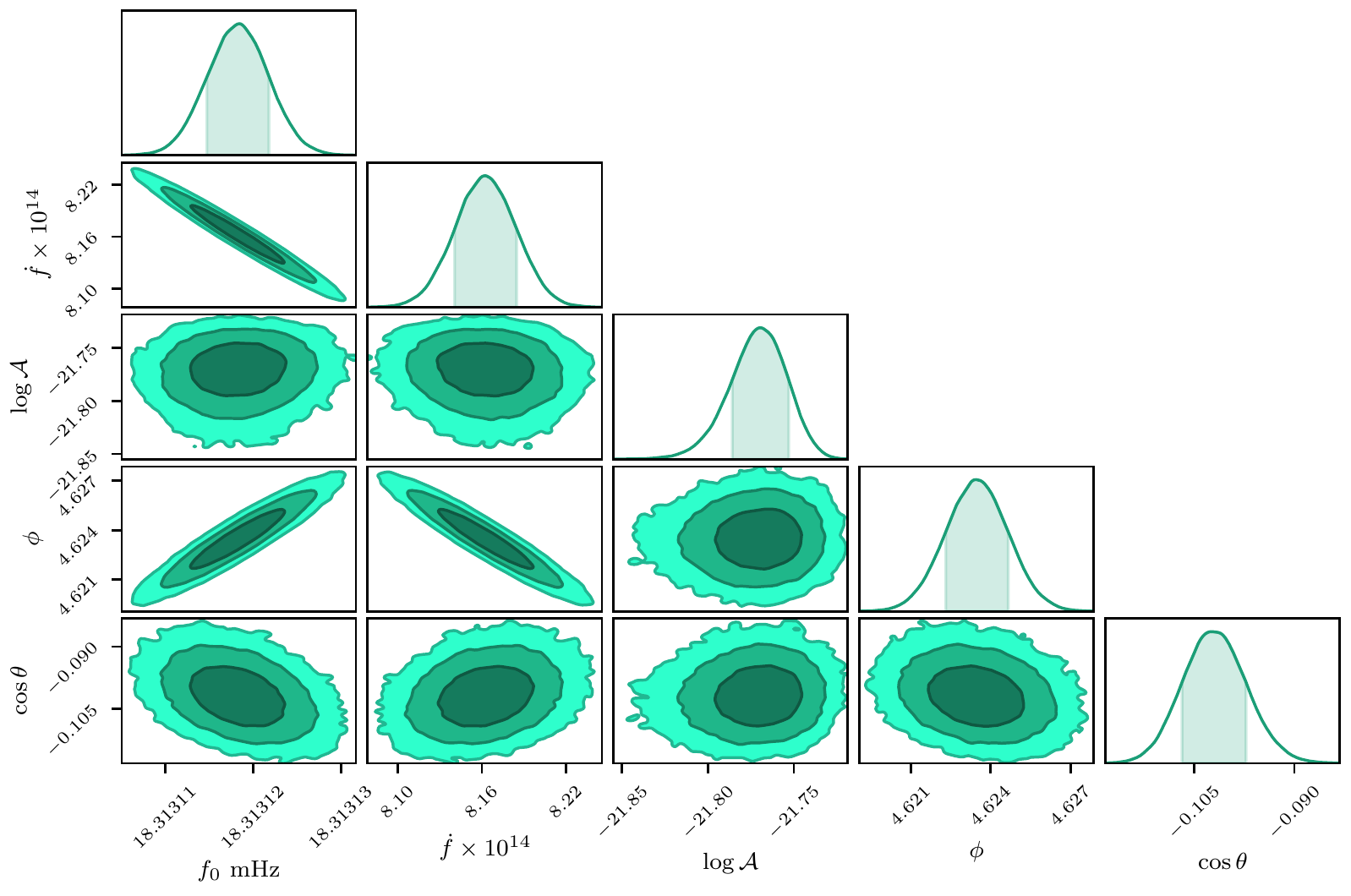} 
\caption{\label{fig:money_plot} Demonstration of the algorithm on a single, isolated, high frequency source. The top left panel shows the power spectrum of the data (black) after 1 year of observations, the posterior distribution of the residual (light blue), and the inferred noise level (light green).  The residual and noise level are plotted as the median with 50\% and 90\% credible intervals. The bottom left panel shows the reconstructed signal waveform posterior (green) identified by the median frequency of the posterior distribution, $\f^{\rm med} = 0.0183131182\ \rm{Hz}$. The right panel is a corner plot showing the marginalized posterior distributions of select parameters likely of most interest to the research community, including the frequency $\f$, frequency derivative $\dot{f}$, amplitude $\mathcal{A}$, and sky location ($\theta,\phi$).}
\end{figure*}

\section{Previous work}
Compared to other GW sources, UCBs are  simple to model.  
When in the LISA band, the binary is widely separated and the stars' velocities are small compared to the speed of light $c$. 
Therefore the waveforms are well predicted using only leading order terms for the orbital dynamics of the binary~\cite{Peters_1963} and appear as nearly monochromatic (constant frequency) sources. Accurate template waveforms are computed at low computational cost using a fast/slow decomposition of the waveform convolved with the instrument response~\cite{Cornish:2007if}. 

The UCB population is nevertheless a challenging source for LISA analysis due to the sheer number of sources expected to be in the measurement band, rather than the complication of detecting and characterizing individual systems. 
Each source is well-modeled by $\mathcal{O}(10)$ parameters and over $10^4$ sources are expected to be individually resolvable by LISA, resulting in a ${\sim}10^5$ parameter model and thus ruling out any brute-force grid-based method. 
Compounding the challenge is the fact that the GW signals, though narrow-band, are densely packed within the LISA measurement band to the extent that sources are overlapping. 
As a consequence, a hierarchichal/iterative scheme where bright sources are removed and the data is reanalyzed produces biased parameter estimation and poorer detection efficiency: Each iteration leaves behind some residual due to imperfect subtraction, and enough iterations are required for the residuals to build up to the point where they limit further analysis~\cite{gClean}. 
It was determined in the early 2000s that stochastic sampling algorithms performing a global fit to the resolvable binaries, while simultaneously fitting a model for the residual confusion or instrument noise and using Bayesian model selection to optimize the number of detectable sources, provided an effective approach. 

The first full-scale demonstration of a galactic binary analysis was put forward by Crowder and Cornish~\cite{Cornish:2005qw,Crowder:2006eu} with the Blocked Annealed Metropolis (BAM) Algorithm.  
The BAM Algorithm started from the full multi-year data set provided by the Mock LISA Data Challenges (MLDCs)~\cite{Babak_2008}.
Because the sources are narrow-band compared to the full measurement band of the detector, the search was conducted independently on sub-regions in frequency.
The analysis region in each segment was buffered by additional frequency bins that overlapped with neighboring segments.
The noise spectrum was artificially increased over the buffer frequencies to suppress signal power from sources in neighboring bands which spread into the analysis window.

The template waveforms were computed in the time domain, Fourier transformed, and tested against the frequency domain data. 
In accordance to the MLDC simulations, the waveform model did not include the intrinsic frequency evolution of the binaries, and the frequency-dependent detector noise level was assumed to be known \emph{a priori}. 
The BAM analysis was a quasi-Bayesian approach, using a generalized multi-source $\mathcal{F}$ statistic likelihood that maximized, rather than marginalized, over four of the extrinsic parameters of each waveform. 
Model parameters used flat priors except for the sky location which was derived from an analytic model for the spatial distribution of binaries in the galaxy, projected onto the sky as viewed by LISA.
To improve the convergence of the algorithm, particularly for high-$\SNR$ signals, the sampler used simulated annealing~\cite{Kirkpatrick671} during the burn-in phase.  

To sample from the likelihood function, BAM employed a custom Markov Chain Monte Carlo (MCMC) algorithm with a mixture of proposal distributions including uniform draws from the prior, jumps along eigenvectors of the Fisher information matrix for a given source, and localized uniform jumps over a range scaled by the estimated parameter errors. The BAM Algorithm made use of domain knowledge by explicitly proposing jumps by the modulation frequency $f \rightarrow f \pm 1/{\rm yr}$ to explore sidebands of the signal imparted by LISA's orbital motion. 

To determine the number of detectable sources, BAM employed an approximate Bayesian model selection criteria, where models of increasing dimension (i.e., number of detectable sources) were hierarchically evaluated, starting with a single source in each analysis segment and  progressively adding additional sources to the fit. 
The different dimension models were ranked using the Laplace approximation to the Bayesian evidence~\cite{Jeffreys61}. 
The stopping criteria for the model exploration was met when the approximated model evidence reached a maximum.

In response to the next generation of MLDCs, Littenberg~\cite{Littenberg:2011zg} extended the BAM Algorithm in several key ways, but maintained the original concept of analyzing independent segments with attention paid to the segment boundaries to avoid edge effects. 
The primary advancement of this generation of the search pipeline was the use of replica exchange between chains of different temperatures (parallel tempering)~\cite{PhysRevLett.57.2607} and marginalizing over the number of sources in the data (as opposed to hierarchically stepping through models) using a Reversible Jump MCMC (RJMCMC)~\cite{doi:10.1093/biomet/82.4.711} to identify the range of plausible models.
To guard against potentially poor mixing of the RJMCMC a dedicated fixed-dimension follow-up analysis with Bayesian evidence computed via thermodynamic integration~\cite{Goggans_2004} was used for the final model selection determination. 
The algorithm continued using the $\mathcal{F}$ statistic likelihood and simulated annealing during burn in (the ``search phase'') but switched to the full likelihood, sampling over all model parameters, during the parameter estimation and model selection phase of the analysis. 
The algorithm additionally made use of the burn-in by building proposal distributions from the biased samples derived during the non-Markovian search phase using a naive binning of the model parameters. 
The algorithm included a parameterized noise model, by fitting coefficients to the expected noise power spectral density (proportional to the variance of the noise). 
The waveform model included frequency evolution, and was computed directly in the Fourier domain using the fast-slow decomposition described in~\cite{CornLitt07}.

Experience gained from the noise modeling and trans-dimensional algorithms originally applied to the LISA galactic binary problem permeated into analyses of ground-based GW data from the LIGO-Virgo detectors.
For spectral estimation the \BayesLine algorithm uses a two-component phenomenological model to measure the frequency-dependent variance of the detector noise~\cite{Littenberg:2015}, while the \BayesWave algorithm uses a linear combination of wavelets to fit short-duration non-Gaussian features in the data~\cite{Cornish:2014kda}. 
The wavelet model in each detector is independent when fitting noise transients, and is coherent across the network when fitting GW models. 
The Bayes factor between the coherent and incoherent models is used as a detection statistic as part of a hierarchichal search pipeline~\cite{PhysRevD.93.022002}.
The number of wavelets, and components to the noise model, are all determined with an RJMCMC algorithm.
The large volume of data, number of event candidates, and thorough measurement of search backgrounds motivated development of global proposal distributions to improve convergence times of the samplers.
The \BayesWave and \BayesLine models were both inspired by the previous work on the galactic binary problem, with the wavelets substituting the UCB waveforms and the \BayesLine model replacing the confusion noise fits.
Completing the feedback loop, lessons learned from the development and deployment of the methods on the LIGO-Virgo data have formed part of the foundation in this work, particularly through global proposal distributions, numerical methods for reducing computational time of likelihood evaluations, and infrastructure for deploying the pipeline on distributed computing resources.

\section{A New Hope} 

The new UCB algorithm incorporates many of the features from the earlier efforts, but improves on them in several ways. 
The biggest change is the adoption of a time evolving strategy, which reflects the reality of the data collection.
Analyzing the data as it is acquired also eliminates dedicated algorithm tuning choices for dealing with very loud sources.
When new data are acquired the analysis starts on the residual after the bright sources identified previously are removed from the data. 
In each analysis segment, the removed sources are added back into the data before the RJMCMC begins sampling.  
This eliminates the problem of having power leakage between analysis segments, and the resultant noise model manipulation to suppress the model from being biased by edge effects in each segment.
The time-evolving analysis is also naturally ``annealed'' as the $\SNR$ of sources builds slowly over time. 

Other significant changes include improvements to the RJMCMC implementation with the addition of global proposal distributions which eliminate the need for a separate, non-Markovain, search phase or the fixed-dimension follow-up analysis for evidence calculation--the model selection is now robustly handled by the RJMCMC itself as originally intended. 

For the first time in the context of our UCB work, we have also considered how to distill the unwieldy output from the RJMCMC to more readily useable, higher-level, data products which is how the majority of the research community will interact with the LISA observations. 

The code described in this work is open source and available under the GNU/GPL v2 license~\cite{littenberg_tyson_2020_3756199}.


\section{Model and Implementation}

Bayesian inference requires the specification of a likelihood function and prior probability distributions for the model components. The implementation of the analysis employs stochastic sampling techniques, in our case the trans-dimensional Reversible Jump Markov Chain Monte Carlo (RJMCMC)~\cite{doi:10.1093/biomet/82.4.711} algorithm with replica exchange~\cite{PhysRevLett.57.2607}, to approximate the high dimensional integrals that define the marginalized posterior distributions. As with all MCMC algorithms, the choice of proposal distributions is critical to the performance. Here we detail the model and the implementation, hopefully in sufficient detail for the analysis to be repeated by others. 

\subsection{Likelihood function}

The LISA science analysis can be carried out using any complete collection of Time Delay Interferometry (TDI) channels~\cite{Prince:2002hp, Adams:2010vc}. For example, we could use the set of Michelson-type channels $I=\{X,Y,Z\}$, or any linear combination thereof. Schematically we can write $\data_I =  {\bf h}_I + {\bf n}_I$, where ${\bf}h_I$ is the response of the $I^{\rm th}$ channel to all the gravitational wave signals in the Universe, and ${\bf n}_I$ is the combination of all the noise sources impacting that channel. Here the ``noise'' will include gravitational wave signals that are individually too quiet to extract from the data. The goal of the analysis is to reconstruct the detectable gravitational wave signal using a signal model ${\bf h}_I$ such that the residual ${\bf r}_I = \data_I - {\bf h}_I$ is consistent with the noise model.  For Gaussian noise the likelihood is written as:
\begin{equation}
p(\data | {\bf h}) = \frac{1}{(2\pi \, \det{\bf C})^{1/2}} \, e^{- \frac{1}{2}(d_{Ik} - h_{Ik}) C^{-1}_{(I k)(J m)} (d_{Jm} - h_{Jm})}\, ,
\end{equation}
where ${\bf C}$ is the noise correlation matrix, and the implicit sum over indicies spans the TDI channels $I=\{X,Y,Z\}$ and the data samples $k$. If the data are stationary, then the noise correlation matrix is partially diagonalized by moving to the frequency domain: $C_{(I k)(J m)} = S_{IJ}(f_k) \delta_{km}$, where $S_{IJ}(f)$ is the cross-power spectral density between channels $I,J$~\cite{Adams:2010vc}. The cross-spectral density matrix is diagonalized by performing a linear transformation in the space of TDI variables. If the noise levels are are equal on each spacecraft, this leads to the $I'=\{A,E,T\}$ variables~\cite{Prince:2002hp} via the mapping
\begin{equation}
\left[ \begin{array}{c} A \\  \\ E \\  \\ T \end{array} \right] = \begin{bmatrix} \frac{2}{3} & -\frac{1}{3} & -\frac{1}{3} \\  && \\ 0 & - \frac{1}{\sqrt{3}}&  \frac{1}{\sqrt{3}} \\ && \\ \frac{1}{3} &  \frac{1}{3} &  \frac{1}{3}  \end{bmatrix} \left[ \begin{array}{c} X \\  \\Y \\  \\Z \end{array} \right]
\end{equation}
In practice, the noise levels in each spacecraft will not be equal, and the $\{A,E,T\}$ variables will not diagonalize the noise correlation matrix~\cite{Adams:2010vc}. However, $\{A,E,T\}$ serve another purpose as they diagonalize the gravitational wave polarization response of the detector for signals with frequencies $f < f_* = 1/(2 \pi L) \simeq 19.1 \; {\rm mHz}$, such that $A \sim h_+$, $E \sim h_\times$ and $T \sim h_\odot$. Since the breathing mode $h_\odot$ vanishes in general relativity, the gravitational wave response of the $T$ channel is highly suppressed for $f < f_*$, making the $T$ channel particularly valuable for noise characterization and the detection of stochastic backgrounds~\cite{Tinto:2001ii,Hogan:2001jn} and un-modeled signals~\cite{travis4}. 

Full expressions for the instrument noise contributions to the cross spectra $S_{IJ}(f)$ are given in Ref.~\cite{Adams:2010vc}. Added to these expressions will be contributions from the ``confusion noise'' from the millions of signals that are too quiet to detect individually.  
The confusion noise will add to the overall noise as well as introduce off-diagonal terms in the frequency domain noise correlation matrix ${\bf C}$, as the confusion noise is inherently non-stationary with periodic amplitude modulations imparted by LISA's orbital motion~\cite{PhysRevD.69.123005}.

For now we have made a number of simplifying assumptions that will be relaxed in future work: We ignore the non-stationarity of the noise and assume that the noise correlation matrix is diagonal in the frequency domain; In addition, since we are mostly interested in signals with frequencies well below the transfer frequency $f_* \simeq 19.1 \; {\rm mHz}$, we only use the $A$ and $E$ data combinations in the analysis, and we assume that the noise in these channels is uncorrelated; Rather than working with a component level model for the noise, as was done in Ref.~\cite{Adams:2010vc}, we break the analysis up into narrow frequency bands $[f_i, f_i+\Delta f]$ and approximate the noise in each band as an undetermined constant $S_i$. The noise level in each band becomes a parameter to be explored by the RJMCMC algorithm, resulting in a piecewise fit to the instrument noise over the full analysis band.

The signal model $\templatesum(\Params)$ is the superposition of each individual UCB in the model parameterized by $\params$:
\begin{equation}
\templatesum_I(\Params) = \sum_{a=0}^{\Nsources} \template_I(\params_a)
\end{equation}
where $\template_I(\params_a)$ denotes the detector response of the $I^{\rm th}$ data channel to the signal from a galactic binary with parameters $\params_a$. Note that the number of detectable systems, $\Nsources$, is {\it a priori} unknown, and has to be determined from the analysis. Indeed, we will arrive at a probability distribution for $\Nsources$, which implies that there will be no single definitive source catalog.
The individual binary systems are modeled as isolated point masses on slowly evolving quasi-circular orbits neglecting the possibility of orbital eccentricity~\cite{Seto:2001pg}, tides~\cite{2012MNRAS.421..426F} or third bodies~\cite{Robson:2018svj}. The signals are modeled using leading order post-Newtonian waveforms. The instrument response includes finite arm-length effects of the LISA constellation and arbitrary spacecraft orbits, but the TDI prescription currently implemented makes the simplifying assumption that the arm lengths are equal and unchanging with time. Adopting more realistic instrument response functions increases the computational cost but does not change the complexity of the analysis. 

To compute the waveforms, a fast/slow decomposition is employed that allows the waveforms to be modeled efficiently in the frequency domain~\cite{Cornish:2007if}. The basic idea is to use trigonometric identities to re-write the detector response to the signal in the form $h(t) = a(t) \cos(2 \pi f_k t)$ where $f_k = n_k/T_{\rm obs}$, $n_k = {\rm int}[ \f T_{\rm obs}]$, and $\f$ is the gravitational wave frequency of the signal (twice the orbital frequency) at some fiducial reference time. The Fourier transform of $h(t)$ is then $\tilde h(f) =\frac{1}{2} ( \tilde a(f-f_k) +  \tilde a(f+f_k))$. Since $a(t)$,which includes the orbital evolution and time-varying detector response, varies much more slowly than the carrier signal $\tilde h(f) =\frac{1}{2} \tilde a(f-f_k)$, the Fourier transform of $a(t)$ is computed numerically using a lower sample cadence than needed to cover the carrier. A sample cadence of days is usually sufficient. Note that in the original implementation~\cite{Cornish:2007if} the signal was written as $h(t) = a(t) \cos(2 \pi \f t)$, which was less efficient as it required the convolution $\tilde{h} * \tilde{a}$. By mapping the carrier frequency to a multiple of the inverse observation time the Fourier transform of the carrier becomes a pair of delta functions and the convolution becomes the sum of just two terms, one of which effectively vanishes. 

Each binary is parameterized by $\Nparams$ parameters. $\Nparams$ is typically eight, with $\params \rightarrow (\amp, \f, \fdot, \phase, \iota, \psi, \theta, \phi)$, where $\amp$ is the amplitude, $\f$ is the initial frequency, $\fdot$ is the (constant) time derivative of the, $\phase$ is the initial phase, $\iota$ is the inclination of the orbit, $\psi$ the polarization angle and $\theta,\phi$ are the sky location in an ecliptic coordinate system. If the evolution of the binary were purely driven by gravitational wave emission we could replace the parameters $\left\{\amp, \fdot\right\}$ by the chirp mass ${\cal M}$ and luminosity distance $D_L$ via the mapping
\begin{eqnarray}\label{MD}
\fdot &=& \frac{96}{5} \pi^{8/3} {\cal M}^{5/3} \f^{11/3}  \nonumber \\
\amp &=& \frac{2 {\cal M}^{5/3} \pi^{2/3} \f^{2/3}}{D_L} \, .
\end{eqnarray}
We prefer the $\left\{\amp,\fdot\right\}$ parameterization as it is flexible enough to fit systems with non-GW contributions to the orbital dynamics, e.g.  mass transferring systems, and it is better suited to modeling systems where $\fdot$ is poorly constrained (it is better to have just one parameter filling its prior range than two). For binaries with unambiguously positive $\fdot$, and assuming GW-dominated evolution of the orbit, we resample the posteriors to $\cal M$ and $\D_L$ in post-processing~\cite{Littenberg_2019}.

We also have optional settings to increase $\Nparams$ by including the second derivative of the frequency~\cite{Littenberg_2019} in which case the frequency derivative is no longer constant, so the parameter $\fdot\rightarrow\fdot_0$ is fixed at the same fiducial time as $\f$ and $\phase$.
Additional, optional changes to the source parameterization includes holding an arbitrary number of parameters fixed at input values determined, for example, by EM observations~\cite{Littenberg_2019b}, or to include parameters which use the UCBs as phase/amplitude standards for self-calibration of the data~\cite{Littenberg_2018}.

\subsection{Prior distributions}

The model parameters are given by the $N_n$ noise levels for each frequency band $S_i$ and the collection of $\Nsources\times\Nparams$ signal parameters $\Lambda$. 
The number of noise parameters $N_n$ is fixed by our choice of bandwidth $\Delta f$ and the frequency range we wish to cover in the analysis. In the current configuration of the pipeline we use analysis windows with $\Delta f \sim\mathcal{O}(\mu\rm{Hz})$ in width resulting in $N_n=\mathcal{O}(10^4)$ noise parameters to cover the full measurement band of the mission. 
We use a uniform prior range $S_I \in [10^{-1} S_I(f_i), 10^{2} S_I(f_i)]$ where $S_I(f_i)$ is the theoretical value for the noise level of data channel $I$ used to generate the data. 
In practice the prior ranges on the noise will be set using information from the commissioning phase of the mission.

The total number of detectable signals $\Nsources$ per frequency band are unknown.
We use a uniform prior covering the range $\Nsources \in U[0,30]$. 
For the individual source parameters we used uniform priors on the initial phase $\phase \in [0,2\pi]$ and polarization angle $\psi \in [0,\pi]$, and a uniform prior on the cosine of the inclination $\cos \iota \in [-1,1]$. 
In each analysis window the initial frequency $\f$ was taken to have a uniform prior covering the range $[f_i, f_i+\Delta f]$.

The allowed range of the frequency derivative is informed by population synthesis models which provide information on the mass and frequency distribution of galactic binaries~\cite{Toonen_2012}. 
While the expression for the frequency derivative is only valid for isolated point masses, the balancing of accretion torques and gravitational wave emission in mass-transferring AM CVn type systems is thought to lead to a similar magnitude for the frequency derivative, but with the sign reversed~\cite{Nelemans2010}. 
Using these considerations as input, we adopt a uniform prior on $\fdot$ in each frequency band that covers the range $\fdot =[ - 5\times 10^{-6} f_i^{13/3}, 8\times 10^{8} f_i^{11/3}]$.  

\begin{figure}[htp]
\includegraphics[width=0.5\textwidth]{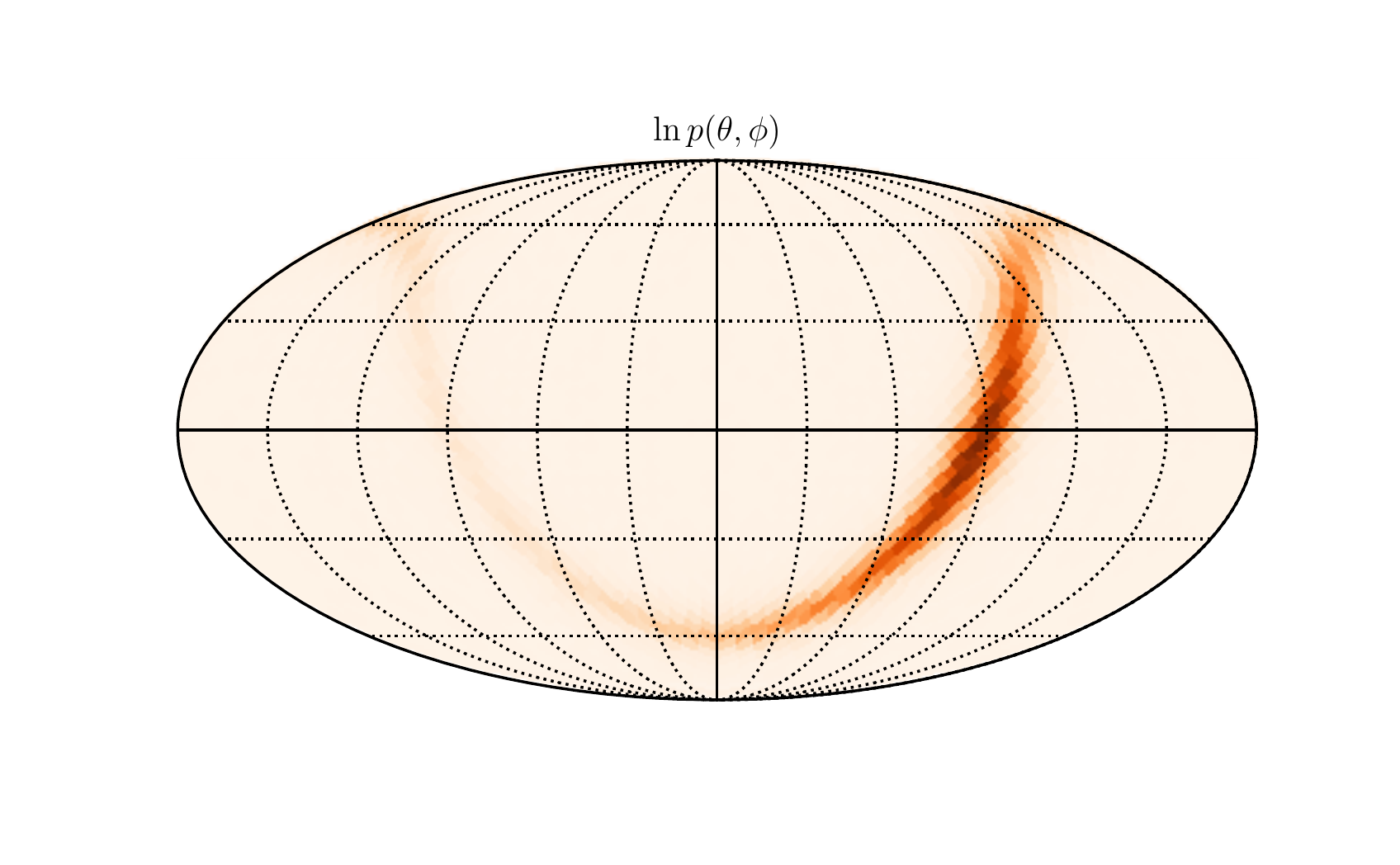} 
\caption{\label{fig:skyprior} The sky prior plotted in ecliptic coordinates. The color scale is logarithmic prior density $\ln p(\theta,\phi)$.}
\end{figure}


For RJMCMC algorithms with scale parameters--in our case the amplitude--the choice of prior influences both the recovery of those parameters as well as on the model posterior.
For example, a simple uniform prior between $U[0,\amp_{\rm max}]$ will support including low-amplitude sources in the model.
Adding a source to the model with $\SNR\sim0$ will not degrade the likelihood, and the remaining model parameters will sample their prior such that the so-called ``Occam penalty'' from including extra (constrained) parameters is small. 
The need to derive an amplitude prior that results in model posteriors as we intuitively expect--namely that templates are included in the model predominantly when there is a detectable source for them to fit--and does not bias the recovery of the amplitude parameter was addressed in the \BayesWave algorithm~\cite{Cornish:2014kda}.
There the prior on the amplitudes had to be considered to suppress large numbers of low-amplitude wavelets saturating the model prior.
The solution was to evaluate the prior not on the amplitudes themselves, but on the {\SNR} of the wavelet. The prior was tuned to go to 0 at low \SNR, peak in the regime where most wavelets were expected to appear in the model (near the ``detection'' threshold), and taper off at high \SNR.
We adopt that approach for the UCB model as follows.

Up to geometrical factors of order unity, the $\SNR$ of a galactic binary $\rho$ is related to the amplitude via the linear mapping
\begin{equation}
\rho = \frac{\amp}{2} \left( \frac{T_{\rm obs}\sin^2(\f/f_*)}{S_A(\f)}\right)^{1/2} \, .
\end{equation}
The prior on the amplitude is then mapped from a prior on $\rho$ of the form
\begin{equation}
p(\rho) = \frac{3 \rho}{4 {\rho}_*^2 (1 + {\rho}/(4{\rho}_*))^5}
\end{equation}
which peaks at $\rho=\rho_*$ and falls off as $\rho^{-4}$ for large $\rho$. Because most detections will be close to the detection threshold we set $\rho_* = 10$. For bright sources the likelihood, which scales as $e^{\rho^2}$, overwhelms the prior, and there is little influence in the the recovered amplitudes from our choice of prior. 

For the sky location the pipeline has support for two options: Either the model can use a uniform priors on the sky or a prior weighted towards the sources being distributed in the galaxy according to an analytic model for its overall shape. 
As currently implemented we use a simple bulge-plus-disk model for the stellar distribution of the form
\begin{equation}
\varrho = \varrho_0 \left[ \alpha e^{-r^2/R_b^2} + (1-\alpha) e^{-u/Rd} \rm{sech}^2\left(\frac{z}{Z_d}\right)\right].
\end{equation}
Here $r^2 = x^2+y^2 +z^2$ and $u^2=x^2+y^2$, and $x,y,z$ are a set of Cartesian coordinates with origin at the center of the galaxy and the $z$ axis orthogonal to the galactic plane.
The parameters are the overall density scaling $\varrho_0$, bulge fraction $\alpha$, bulge radius $R_b$, disk radius $R_d$ and disk scale height $Z_d$. Ideally we would make these quantities hyper-parameters in a hierarchical Bayesian scheme~\cite{Adams:2012qw}, but for now we have fixed them to the fiducial values $\alpha=0.25$, $R_b=0.8$ kpc, $R_d=2.5$ kpc,  and $Z_b=0.4$ kpc and $\varrho_0$ determined by numerically normalizing the distribution, . LISA views the galaxy from a location that is offset from the galactic center by an amount $R_G$ in the $x$-direction, and use ecliptic coordinates to define the sky locations. This necessitates that we apply a translation and rotation to the original galactic coordinates. We then compute the density $\varrho(\theta,\phi)$ in the new coordinate system and normalize the density on the sky to unity for use as a prior. In order to ensure full sky coverage we rescale the normalized density by a factor of $(1-\beta)$ and add to it a uniform sky distribution that has total probability $\beta$. Figure~\ref{fig:skyprior} shows the sky prior for the choice $\beta = 0.1$.

\subsection{Trans-dimensional MCMC}

Trans-dimension modeling is a powerful technique that simultaneously explores the range of plausible models for the data as well as the parameters of each candidate model. The trans-dimensional approach is particularly valuable in situations where it is unclear how many components should be included in the model and there is a danger of either over- or under-fitting the data.
Trans-dimensional modeling allows us to explore a wide class of models in keeping with our motto ``model everything and let the data sort it out''~\cite{Cornish:2014kda}. 
While fixed dimension (signal model) sampling techniques have thus far proven sufficient for LIGO-Virgo analyses of isolated events, we see no alternative to using trans-dimensional algorithms for the multi-source fitting required for LISA data analysis.

Trans-dimensional MCMC algorithms are really no different from ordinary MCMC algorithms. They simply operate on an extended parameter space that is written in terms of a model indicator parameter $k$ and the associated parameter vector $\vec{\theta}_k$. It is worth noting that the number of models can be vast. For example, suppose we were addressing the full LISA data analysis problem using a model that included up to $N_{\rm UCB}\sim 10^5$ galactic binaries, $N_{\rm BH} \sim 10^3$ supermassive black holes, $N_{\rm EMRI}\sim 10^3$ extreme mass ratio inspirals and $N_{\rm n}\sim 10^3$ parameters in the noise model. Since the number of parameters for each model component are not fixed, the total number of possible models is the {\em product}, not the sum, of the number of possible sub-components, resulting in $\sim 10^{14}$ possible models in this instance. 
The advantage of the RJMCMC method is that it is not necessary to enumerate or sample from all possible models but, rather, to have the {\em possibility} of visiting the complete range of models. This is in contrast to the product space approach~\cite{10.2307/2346151}, which requires that all models be enumerated and explored while most of the computing effort is spent exploring models that have little or no support. 
Just as an ordinary MCMC spends the majority of its time exploring the regions of parameter space with high posterior density, the RJMCMC algorithm spends most of the time exploring the most favorable models.
 
Our goal is to compute the joint posterior of model $k$ and parameters $\gparams_k$
\begin{equation}
p(k, \gparams_k | \data) = \frac{ p(\data | k, \gparams_k)  p(k, \gparams_k)}{ p(\data)}
\end{equation}
which is factored as
\begin{equation}
p(k, \gparams_k | \data) =  p(k | \data) p(\gparams_k | k, \data) \, ,
\end{equation}
where $p(k|\data)$ is the posterior on the model probabilities and $p(\gparams_k | k, \data)$ is the usual parameter posterior distribution for model $k$. The quantity $O_{ij} = p(i | \data)/p(j | \data)$ is the odds ratio between models $i,j$. The RJMCMC algorithm generates samples from the joint posterior distribution $p(k, \vec{\theta}_k | \data)$ by developing a Markov Chain via proposing transitions from state $\{k, \gparams_k\}$ to state $\{l,\gparams_l\}$ using a proposal distribution $q(\{k, \gparams_k\}, \{l,\gparams_l\})$. Transitions are accepted with probability $\alpha = \min \left\{ 1, H_{l\rightarrow k} \right\}$ with the Hastings Ratio
\begin{equation}\label{rjmcmc}
H_{l\rightarrow k} = \frac{p(\data | k, \gparams_k)}{p(\data | l, \gparams_l)} \, \frac{p(k, \gparams_k)}{p(l, \gparams_l)} \, \frac{q(\{k, \gparams_k\}, \{l,\gparams_l\})}{q(\{l,\gparams_l\}, \{k, \gparams_k\} )}. 
\end{equation}
Proposals are usually separated into within-model moves, where $k=l$ and only the model parameters $\gparams_k$ are updated, and between-model moves where both the model indicator $l$ and the model parameters $\gparams_l$ are updated. Written in the form of Eq.~\ref{rjmcmc} the RJMCMC algorithm is no different than the usual Metropolis-Hastings algorithm. In practice the implementation is complicated by the need to match dimensions between the model states, which introduces a Jacobian determinant of the mapping function~\cite{doi:10.1093/biomet/82.4.711}. This can all become very confusing, and may explain the slow adoption of trans-dimensional modeling in the gravitational wave community. Thankfully the models we consider are {\em nested}, such that the transition from state $k$ to $l$ involves the addition or removal of a model component. In the case of nested models the mapping function is a linear addition or subtraction of parameters, and the Jacobian is simply the ratio of the prior volumes~\cite{doi:10.1111/j.1365-246X.2006.03155.x}. For example, the Hasting ratio for adding a single UCB source with parameters $\params_{k+1}$ to the current state of the model already using $k$ templates (with joint parameters $\Params_k$) is
\begin{equation}
    H_{k\rightarrow k+1} = \frac{p(\data |\Params_k, \params_{k+1}) p(\params_{k+1}) } {p(\data |\Params_k) q(\params_{k+1})}
\end{equation}
where $q(\params_{k+1})$ is the proposal distribution that generated the new source parameters, and we assume for the reverse move ($k+1\rightarrow k$) that existing sources are selected for removal with uniform probability. 

The efficiency of any MCMC algorithm depends critically on the choice of proposal distributions. The necessity for finding good proposal distributions is even more acute for the trans-dimensional moves of a RJMCMC algorithm. 
In the UCB pipeline, an increase in dimension comes about when a new waveform template is added to the solution. 
For such a move to be accepted the parameters for the new source must land sufficiently close to the true parameters of some signal for the transition to be accepted. Arbitrarily choosing the $\Nparams$ parameters that define a signal has low probability of improving the likelihood enough for the transition to be accepted. 
The strategy we have adopted to improve the efficiency, which is explicitly detailed in the following section, is to identify promising regions of parameter space in pre-processing, in effect producing coarse global maps of the likelihood function, and using these maps as proposal distributions. 
The global proposals are also effective at promoting exploration of the multiple posterior modes that are a common feature of GW parameter spaces for single sources.

To further aid in mixing we use replica exchange (also know as parallel tempering). 
Parallel tempering uses a collection of chains to explore models with the modified likelihood $p(\data | \Params, \beta) = p(\data | \Params)^{\beta}$, where $\beta\in[0,1]$ is an inverse ``temperature''. Chains with high temperatures (low $\beta$) explore a flattened likelihood landscape and move more easily between posterior modes, while chains with lower temperature sample the likelihood around candidate sources and map out the peaks in more detail. Only those chains with $\beta=1$ provide samples from the target posterior.
A collection of chains at different temperatures are run in parallel, and information is passed up and down the temperature ladder by proposing parameter swaps, which are accepted with probability $\alpha = \min\left\{ 1,H_{i\leftrightarrow j}\right\}$ and 
\begin{equation}\label{ptmcmc}
H_{i\leftrightarrow j} = \frac{ p(\data | i, \Params_i,\beta_i) \, p(\data | j, \Params_j,\beta_j)}{  p(\data | i, \Params_i,\beta_j) \, p(\data | j, \Params_j,\beta_i) } \, .
\end{equation}
Here we are proposing to swap the parameters of the model $\{i, \Params_i\}$ at inverse temperature $\beta_i$ with the model $\{j, \Params_j\}$ at inverse temperature $\beta_j$. Note that if $\beta_i=\beta_j$ the swap is always accepted. Models with higher temperatures typically have lower likelihoods. If the likelihoods of the two models are very different the Hastings Ratio $H_{i\leftrightarrow j}$ will be small. 
We only propose exchanges between chains that are near one another in temperature. 

Choosing the temperature ladder so that chain swaps are readily accepted is a challenge. The situation we need to avoid is a break in the chain, where a collection of hotter chains decouples from the colder chains such that no transitions occur between the two groups. When that happens the effort spent evolving the hot chains is wasted as their findings are never communicated down the temperate ladder to the $\beta=1$ chain(s) that accumulate the posterior samples. It is generally more effective to run a large number of chains that are closely spaced in temperature for few iterations than it is to run with fewer chains for longer. We adopt the scheme described in Ref.~\cite{Vousden_2015} where the temperature spacing between chains is adjusted based on acceptance rates of chain swaps, and the degree to which the temperatures adjust based on the acceptance rates, asymptotically approaches zero as the number of chain iterations increases.
Thus the temperature spacing is dynamically adjusting to the rapidly changing model when the sampler is ``burning in'' but settles into a steady-state when the sampler is exploring the posterior.

\subsection{Proposal Distributions}

\begin{figure}[htp]
\includegraphics[clip=true,angle=0,width=0.5\textwidth]{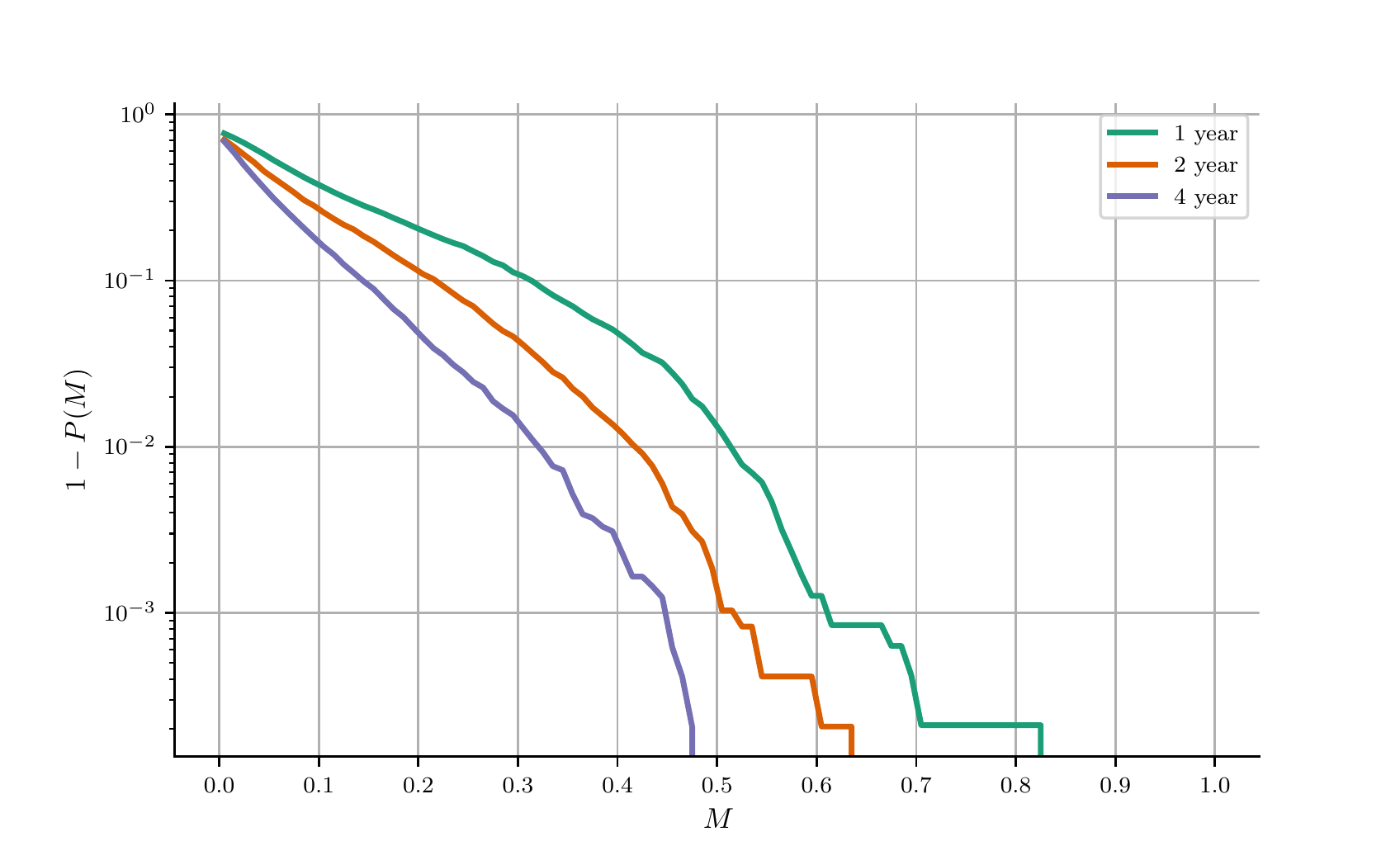} 
\caption{\label{fig:overlap} Survival function of the maximum match between any pair of detectable sources computed using a simulated galactic population of UCBs. For 1 year of observing (green) $\lesssim1\%$ of sources have overlaps greater than 50\%. That fraction is reduced to $0.1\%$ after 2 years (orange), and $0$ after 4 years (purple) as the resolving power of LISA increases.}
\end{figure}

As mentioned previously, the efficiency of a MCMC algorithm is heavily dependent on the design of the proposal distributions.  
This ``tuning'' requirement for an efficient MCMC has led to the development of samplers designed to be more agnostic to the parameter space such as ensemble samplers (e.g.~\cite{Foreman_Mackey_2013}), Hamiltonian Monte Carlo~\cite{betancourt2017conceptual}, etc. 
However, there has been less development of alternatives to sampling transdimensional posteriors and the scale of the LISA UCB problem may be prohibitive to brute-force evaluation of many competing models. 
It is our view that continued innovation in development of custom proposal distributions that leverage the hard-earned domain knowledge is worth the investment.
To that end, we observe that the posterior is the ideal proposal distribution--setting $q(\{i,\Params_i\}, \{j, \Params_j\} ) = p(i, \Params_i | \data)$ we have $H_{i\rightarrow j}=1$, so every proposed move is accepted and the correlation between successive samples can be made arbitrarily small. Of course, if we could produce independent samples from the posterior in advance there would be no need to perform the MCMC, but this observation provides guidance in the design of effective proposal distributions--we seek distributions that are computationally efficient approximations to the posterior distribution, which usually amounts to finding good approximations to the likelihood function. Consider the log likelihood for model $k$ describing $N_k$ galactic binaries, which is written as
\begin{eqnarray} \label{ll}
\ln p(\data | k, \Params_k ) &=& \sum_{i=1}^{N_k} \ln p(\data | \params_i) + \frac{1}{2}(N_k-1)\innerproduct{d}{d} \nonumber \\
&+&\sum_{i>j} \innerproduct{\params_i}{\params_j} \, ,
\end{eqnarray}
where
\begin{equation}
\innerproduct{a}{b} \equiv a_{Im}  C^{-1}_{(I m)(J n)}(\vec{\kappa}) b_{Jn} 
\end{equation}
and we are neglecting terms from the noise parameters. The first term in the expression for the log likelihood in Eq.\ref{ll} is the sum of the individual likelihoods for each source, while the final term describes the correlations between the sources. While accounting for these correlations is crucial to the global analysis, the correlation between any pair of sources is typically quite small, and we ignore them in the interest of finding a computationally efficient approximation to the likelihood to use as a proposal. Figure~\ref{fig:overlap} shows the maximum match between pairs of sources with $\SNR > 7$, using a simulated galactic population and assuming 1, 2, and 4 year observation periods. Here the match, or overlap, is defined as:
\begin{equation}\label{eq:match}
M_{ij} \equiv \frac{\innerproduct{\template(\params_i)}{\template(\params_j)}}{\sqrt{ \innerproduct{\template(\params_i)}{\template(\params_i)} \innerproduct{\template(\params_j)}{\template(\params_j)} }}\, ,
\end{equation}
and we are using the $A,E$ TDI data channels. Less than 1\% of sources have overlaps greater than 50\%, and the fraction diminishes with increased observing time.
Thus we will develop proposals for individual sources and propose updates to their parameters independently of other sources in the model.
The MCMC still marginalizes over the broader parameter space, including the rare but non-zero case of non-negligible covariances between sources, in effect executing a blocked Gibbs sampler where the blocks are individual source's parameters.

\subsubsection{$\mathcal{F}$ statistic Proposal}
We construct a global proposal density using the single source $\mathcal{F}$ statistic to compute the individual likelihoods $\ln p(\data | \params_i)$ maximized over the extrinsic parameters $\amp, \phase, \iota, \psi$. Up to constants that depend on the noise parameters, the maximized log likelihood is equal to
\begin{equation}
{\cal F}(\f,\theta, \phi) = \frac{1}{2} \innerproduct{{\bf g}_i}{{\bf g}_j}^{-1} \innerproduct{\data}{{\bf g}_i} \innerproduct{\data}{{\bf g}_j}  
\end{equation}
where the four filters ${\bf g}_i$ are found by computing waveforms with parameters $\f, \fdot=0, \theta, \phi$, $\amp=2$ and 
\begin{eqnarray} \label{filters}
&& {\bf g}_1 = \template\left(\phase=0, \iota=\frac{\pi}{2},\psi=0\right) \\
&& {\bf g}_2 = \template\left(\phase=\pi, \iota=\frac{\pi}{2},\psi=\frac{\pi}{4}\right) \\
&& {\bf g}_3 = \template\left(\phase=\frac{3\pi}{2}, \iota=\frac{\pi}{2},\psi=0\right) \\
&& {\bf g}_4 = \template\left(\phase=\frac{\pi}{2}, \iota=\frac{\pi}{2},\psi=\frac{\pi}{4}\right) \, .
\end{eqnarray}
The $\mathcal{F}$ statistic proposal is the three dimensional histogram precomputed from the data using a grid in $\f, \theta, \phi$. 
We use a fixed grid spacing governed by what is needed for the best resolved sources which are found in the ecliptic plane (which maximises the doppler modulations imparted by LISAs orbital motion) and at the highest frequencies covered by the analysis. 
The probability density of a cell $(a,b,c)$ of the three-dimensional histogram is ${\cal F}_{a,b,c}$ normalized by the sum of ${\cal F}$ over all cells, and the parameter volume of the cell. 

The optimal spacing of the grid can be estimated from the reduced Fisher information matrix $\gamma_{ij}$, which is found by projecting out the parameters $\amp, \phase, \iota, \psi$ from the full Fisher information matrix $\Gamma_{ij} = \innerproduct{\partial\template/\partial\params_i}{\partial\template/\partial\params_j}$~\cite{Cornish:2005qw}. 
The reduced Fisher matrix is not constant across the parameter space and will naturally reduce the grid size as $\f$ gets larger, and for sky locations near the ecliptic equator compared to those near the poles. The grid spacing will also become finer as the observation time grows. 
These modifications, as well as extending to a 4D grid including $\fdot$, will further improve the efficiency of the proposal and are left for future development.

\begin{figure}[htp]
\begin{center}
\includegraphics[clip=true,width=0.5\textwidth]{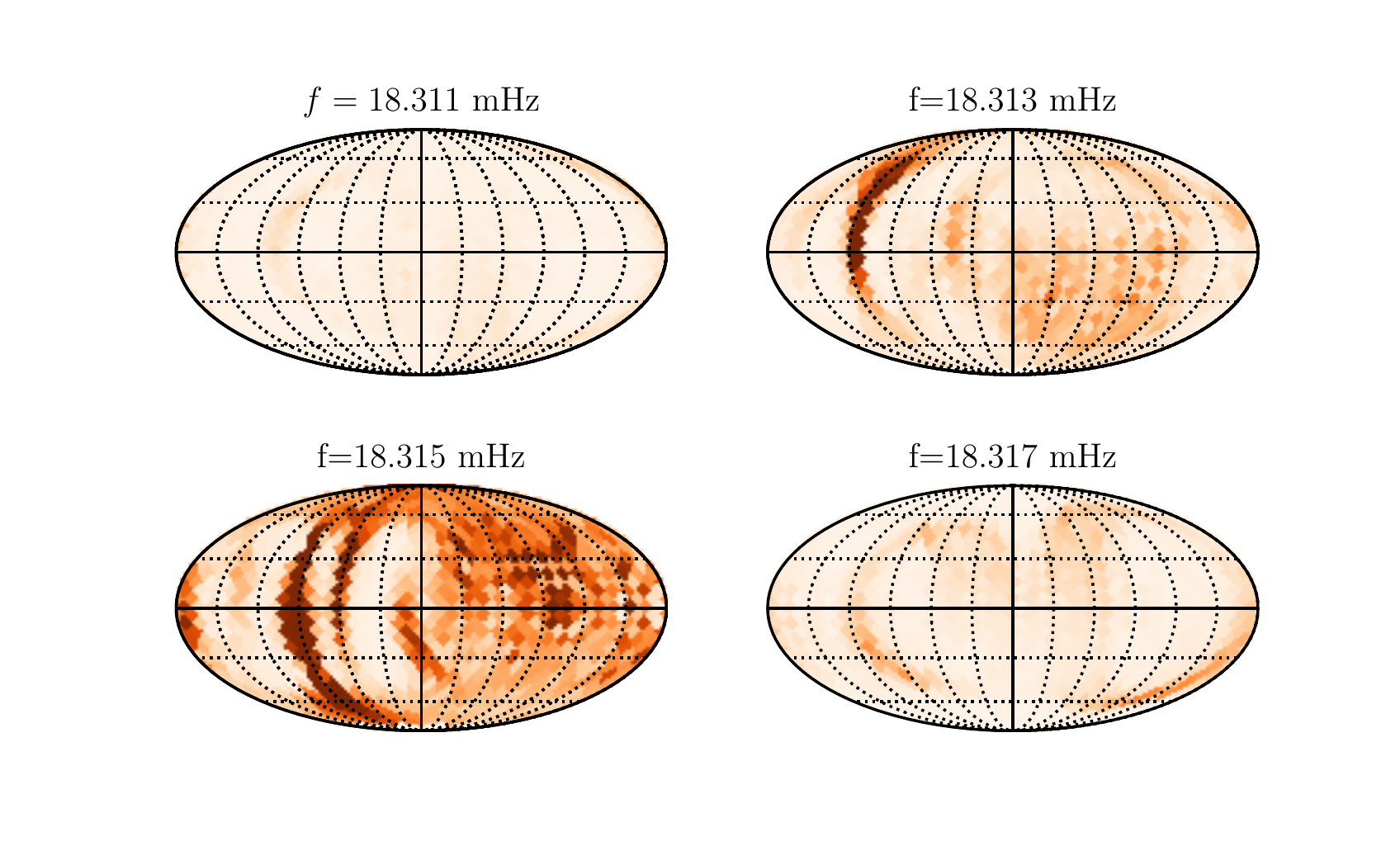} 
\end{center}
\vspace*{-0.2in}
\caption{\label{fig:fstat-proposal} Frequency slices of the multidimensional $\mathcal{F}$ statistic proposal for the same segment of data shown in Fig.~\ref{fig:money_plot}. The color scale is linear in the proposal density, and each panel is on the same scale. The proposal promotes frequencies and sky locations consistent with the signal in the data (top right and bottom left panels) and returns a low-density and diffuse distribution at frequencies consistent with random noise (top left and bottom right panels).}
\end{figure}

\subsubsection{Multi-modal Proposals}

Due to the parameterization of the gravitational wave signals, and the instrument response to those signals, there are known  exact or near degeneracies which appear as distinct modes in the likelihood/posterior distribution. 
While MCMC algorithms are not generically efficient at sampling from multimodal distributions, we have developed dedicated proposal distributions to exploit the predictable multi-modality and improve the chain convergence time.

Due to the annual orbital motion of the LISA constellation, continuous monochromatic sources will have non-zero sidebands at the modulation frequency $f_m = 1/{\rm year}$. 
Sources that are detectable at low $\SNR$ after several years of observation can have likelihood support at multiple modes separated by $f_m$, while for high $\SNR$ sources the secondary modes are subdominant local maxima, challenging to generic MCMC sampling algorithms.
We have adopted a dedicated proposal that updates the UCB initial frequency by $\f \rightarrow \f + n f_m$ where $n$ is drawn from $N[0,1]$ and mapped to the nearest integer.  
The sky location of the source correlates with the frequency through the doppler modulations imparted by the detector's orbital motion, so the proposal alternates between updates to the extrinsic parameters using the Fisher matrix proposal, F-statistic proposal, and draws from the prior. 
A similar proposal was deployed and demonstrated in Refs.~\cite{Crowder:2006eu,Littenberg:2011zg}.

We also take advantage of a linear correlation between the gravitational wave phase $\phase$ and polarization angle $\psi$, and a perfectly degenerate pair of modes over the prior $\psi \in [0,\pi]$ and $\phase \in [0,2\pi]$ by proposing $\{\psi,\phase\}\rightarrow \{\psi \pm \delta/2,\phase \pm \delta\}$ where $\delta \in U[0,2\pi]$ and the sign of the shift in the parameters is random, as the sign of the $\psi/\phase$ correlation depends on the sign of $\cos{\iota}$, i.e. if the stars are orbiting clockwise or counterclockwise as viewed by the observer.

\subsubsection{Posterior-Based Proposals}

The UCBs are continuous sources for LISA and will be detectable from the beginning of operations throughout the lifetime of the mission. 
Our knowledge of the gravitational wave signal from the galaxy will therefore build gradually over time.  
We have designed a proposal distribution to leverage this steady accumulation of information about the galaxy by analyzing the data as it is acquired, and building proposal functions for the MCMC from the archived posterior distributions inferred at each epoch of the analysis.

For a particular narrow-band segment of data, the full posterior is a complicated distribution due to the probabilistically determined number of sources in the data, and their potentially complicated, multimodal structure. 
The posterior is known to us only through the discrete set of samples returned by the MCMC but for use as a proposal must be a continuous function over all of parameter space (as we must be able to evaluate the proposal anywhere in order to maintain detailed balance in the Markov chain).
Therefore some simplifications must be made to convert the discrete samples of the chain into a continuous function. 

In the release of the pipeline accompanying this paper, we select chain samples from the maximum marginalized likelihood (i.e. highest evidence) model at the current epoch to build the proposals used in the subsequent analysis when more data are available..
We post-process the chain samples to cluster those that are fitting discretely identified sources, and to filter out samples from the prior or from weaker candidate sources that don't meet our threshold for inclusion in the source catalog. The post-production analysis is described in Sec.~\ref{sec:catalog}.

Each source $i$ identified in the post-production step will have at least two modes, because of the degeneracy in the $\psi-\phase$ plane.  For each mode $n$, we compute the vector of parameter means $\bar\params_{i,n}$ from the one-dimensional marginalized posteriors, the full $\Nparams\times\Nparams$ covariance matrix $\mathbf{C}_{i,n}$ from the chain samples, and the relative weighting $\alpha_{i,n}$ which is the number of samples in the mode normalized by the total number of samples used to build the proposal. 

The proposal is evaluated for arbitrary parameters $\params$ as
\begin{equation}
p(\params) = \sum_{i=0}^{i<I} \sum_{n=0}^{i<2} \alpha_{i,n} \frac{  e^{-\frac{1}{2} (\params-\bar\params_{i,n}) \mathbf{C}_{i,n}^{-1} (\params-\bar\params_{i,n})} }{\left((2\pi)^{\Nparams} \det\mathbf{C}_{i,n}\right)^{1/2}} .
\end{equation}

To draw new samples from this distribution, we first select which mode by rejection sampling on $\alpha_{i,n}$, and then draw new parameters $\params$ via:
\begin{equation}
\params = \bar\params_{i,n}  + \mathbf{L}_{i,n}\mathbf{n}
\end{equation}
where $\mathbf{n}$ is an $\Nparams$-dimension vector of draws from a zero-mean unit-variance Gaussian, and $\mathbf{L}_{i,n}$ is the LU decomposition of $\mathbf{C}_{i,n}$.

\begin{figure}[htp]
\begin{center}
\includegraphics[clip=true,angle=0,width=0.5\textwidth]{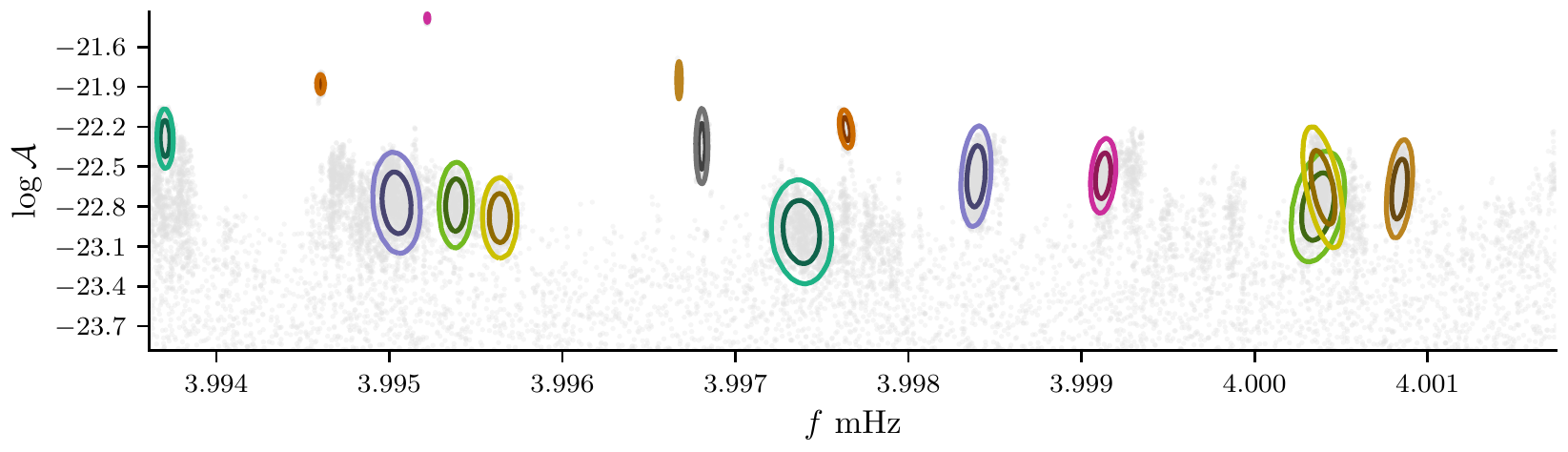} 
\includegraphics[clip=true,angle=0,width=0.5\textwidth]{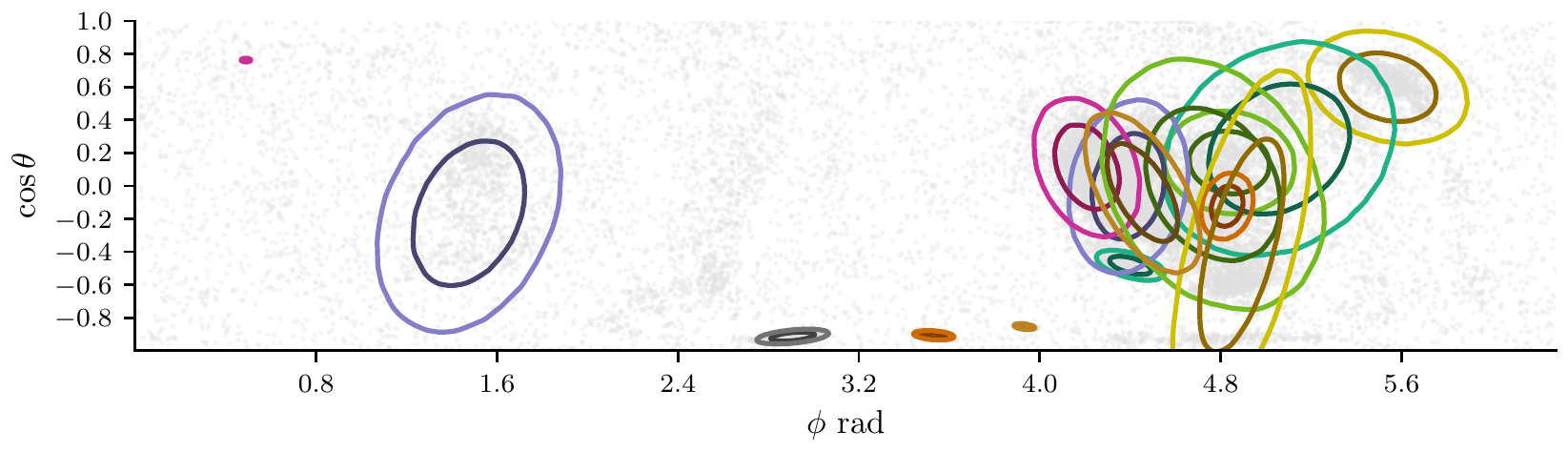} 
\end{center}
\vspace*{-0.2in}
\caption{\label{fig:cov-proposal} Two-dimensional projections of the multi-source covariance matrix proposal produced after analyzing 6 months of simulated data round 4 mHz. The gray scatter plots show all of the chain samples from the analysis which are then filtered and clustered into discrete sources by the catalog production step. The mean parameter values and covariance matrix for each discrete source are computed from the chain samples and used a proposal for the next step of the analysis after more data are acquired. Parameter combinations shown are the frequency-amplitude plane (top panel) and sky location (bottom panel). Ellipses enclose the 1 and $2\sigma$ contours of the covariance matrices, and sources are colored consistently in the top and bottom panels.}
\end{figure}

Fig.~\ref{fig:cov-proposal} shows the 1 and 2$\sigma$ contours of the set of covariance matrices computed from a 6 month observation of simulated LISA data around 4 mHz in two projections of the full posterior: the $\f-\amp$ plane (top) and sky location (bottom). 
Shown in gray is the scatter plot of all chain samples before being filtered by the catalog production step described in the next section. 
The color scheme is consistent between the two panels. 
Note that for well localized (e.g. high amplitude) sources the covariance matrix is a good representation of the posterior, as should be the case since the posterior should trend towards a Gaussian distribution with increased $\SNR$, and will therefore serve as an efficient proposal when new data are acquired.

\begin{figure}[htp]
\begin{center}
\includegraphics[width=0.5\textwidth]{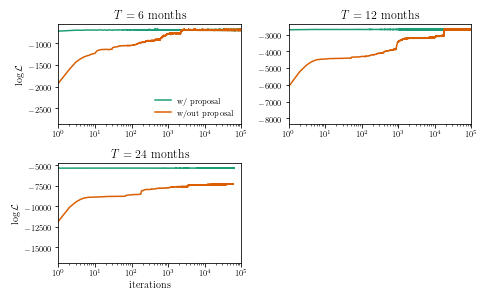}
\vspace*{-0.2in}
\caption{\label{fig:logL} Log-likelihood chains from analyses of the same data as shown in Fig.~\ref{fig:money_plot} run with (teal) and without (orange) the covariance matrix proposal. As the observing time increases, the chain sampling efficiency gained by including the proposal built from previous analyses becomes more significant.}
\end{center}
\end{figure}

Fig.~\ref{fig:logL} shows the log-likelihood of the model as a function of chain step for observations of increasing duration $T$ with (teal) and without (orange) using the covariance matrix proposal built from each intermediate analysis. This demonstration was on the same data from Fig.~\ref{fig:money_plot} containing the type of high-$\f$ and high-$\SNR$ source that proved challenging for the previous RJMCMC algorithm~\cite{Littenberg:2011zg}. 
With the covariance matrix proposal the chain convergence time is orders of magnitude shorter than using the naive sampler, to the point where the $T=24$ month run failed to converge in the number of samples it took the analysis with the covariance matrix proposal to finish.

Using the customized proposals described in this section allows the sampler to robustly mix over model space and explore the parameters of each model supported by the data. 
The pipeline dependably converges without the need for non-Markovian maximization steps as were used in the ``burn in'' phase of our previously published UCB pipelines, and reliably produces results for model selection and parameter estimation analyses simultaneously.

\subsection{Data selection}
While the UCB pipeline is pursuing a global analysis of the data, we leverage the narrow-band nature of the sources to parallelize the processing.
Sources separated by more than their bandwidth--typically less than a few hundred frequency bins--are uncorrelated and can therefore be analyzed independently of one another.

As was done in previous UCB algorithms~\cite{Crowder:2006eu,Littenberg:2011zg}, we divide the full Fourier domain data stream into adjacent segments and process each in parallel, without any exchange of information during the analysis between segments.
To prevent edge effects from templates trying to fit sources outside of the analysis window, each segment is padded on either side in frequency with data amounting to the typical bandwidth of a source, thus overlapping the neighboring segments.  
The MCMC is free to explore the data in the padded region, but during post-production only samples fitting sources in the original analysis window are kept, preventing the same source from being included in the catalog twice.  
Meanwhile, sources within the target analysis region but close to the boundary will not have part of their signal cut off in the likelihood integral.

Unlike in Refs.~\cite{Crowder:2006eu,Littenberg:2011zg}, there is no manipulation of the likelihood or noise model to prevent loud sources outside of analysis region from corrupting the fit.  
Instead, we leverage the time-evolving analysis by ingesting the list of detections from previous epochs of the catalog, forward modeling the sources as they would appear in the current data set and subtracting them from the data.
This will be an imperfect subtraction but is adequate to suppress the signal power in the tails of the source which extend into the adjacent segments and, due to the padding, does not alter the data in the target analysis region. 
In the event that an imperfect subtraction leaves a detectable residual, it will not corrupt the final catalog of detected sources because templates fitting that residual will be in the padded region of the segment and removed in post-processing.  
The downside is merely in the computational efficiency, as poorly-subtracted loud signal with central frequency out of band for the analysis will require several templates co-adding to mitigate the excess power, wasting computing cycles and increasing the burden on the MCMC to produce converged samples.
The effectiveness of the subtraction will improve as the duration of observing time between analyses decreases, and is an area to explore when optimizing the overall cost of the multi-year analysis.

The strategy for mitigating edge effects is prone to failure if the posterior distribution of a source straddles the boundary. The frequency is precisely constrained for any UCB detection so having a source so precariously located is unlikely but nonetheless needs to be guarded against. 
While not yet implemented, we envision checks for sources near the boundaries in post-production to see if posterior samples from different windows should be combined, and/or adaptively choosing where to place the segment boundaries based on the current understanding of source locations from previous epochs of the analysis. There is no requirement on the size or number of analysis windows except that they are much larger than the typical source bandwidth, and the segment boundaries do not need to remain consistent between iterations of the analysis as more data are added.

\begin{figure}[htp]
\begin{center}
    \includegraphics[width=0.5\textwidth]{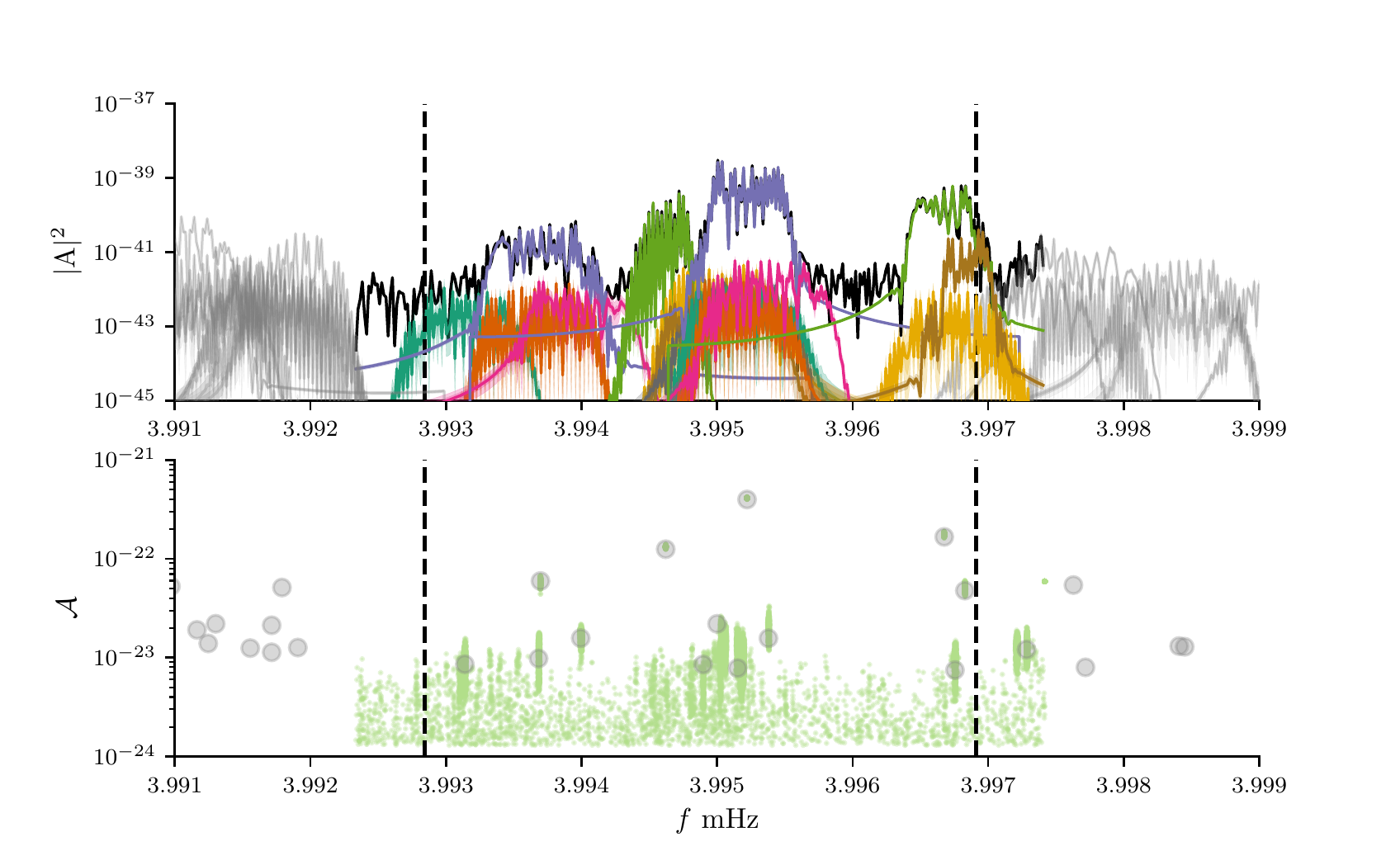}
    \vspace*{-0.2in}
    \caption{\label{fig:padding} Demonstration of data selection and padding procedure. The top panel shows the power spectrum of an example analysis segment in black and the reconstructed waveforms from the analysis in various colors. The vertical dashed lines mark the region of the analysis region where sources will be selected for the catalog. Gray reconstructions are from the analyses of the adjacent segments.  The bottom panel shows the same frequency interval in the $\{f_0,\amp\}$ plane with injected signals marked as gray circles and a scatter plot of the MCMC samples in green.  Note that the chain samples extend into the padded region and fit sources there, but those waveforms are not included in the top-panel's reconstructions}
\end{center}
\end{figure}

Fig.~\ref{fig:padding} demonstrates the data selection and padding procedure by displaying results from the center analysis region of three adjacent windows processed with the time-evolving RJMCMC algorithm. 
The top and bottom panels show the reconstructed waveforms and posterior samples, respectively. The posterior samples extend outside of the analysis region (marked by vertical dashed lines) to fit loud signals in neighboring frequency bins, but are rejected during the catalog production step. The frequency padding ensures that the waveform templates of sources inside of the analysis region are not truncated at the boundary. Sources recovered from the neighboring analyses are marked in gray. Note that there is no conflict between the fit near the boundaries despite their being overlapping sources in this example at the upper frequency boundary.

\section{Catalog Production}\label{sec:catalog}
The output of the RJMCMC algorithm is thousands of random draws from the variable dimension posterior, with each sample containing an equally likely set of parameters \emph{and} number of sources in the model. 
Going from the raw chain samples to inferences about individual detected sources is subtle, as a model using $\Nsources$ templates does not necessarily contain $N$ discrete sources.
For example, the model may be mixing between states where the $\Nsources^{\rm th}$ template is fitting one (or several) weak sources, or sampling from the prior, and such a model could be on similar footing with the  $\Nsources-1$ or $\Nsources+1$ template models purely on the grounds of the evidence calculation.
How then to answer the questions ``How many sources were detected?'' or ``What are the parameters of the detected sources?'' in a way that is robust to the more nuanced cases where the data supports a broad set of models containing several ambiguous candidates?

\subsection{Filtering and Clustering Posterior Samples}

In Ref.~\cite{Littenberg:2011zg}, for the sake of responding to the Mock LISA Data Challenge, post-processing the chains went only as far as selecting the maximum likelihood chain sample from the maximum likelihood model. 
Condensing the rich information in the posterior samples down to single point estimate defeats the purpose of all the MCMC machinery in the first place.  Furthermore, due to the large number of sources being fit simultaneously and the finite number of samples, the maximum likelihood sample within a particular dimension model does not necessarily correspond to the maximum likelihood parameters for each of the many sources in the analysis should they have been fit by the model in isolation.

It was therefore necessary that we begin to seriously consider how to post-process the raw chain samples into a more manageable data product for the sake of producing source catalogs that are easily ingested by end users of the LISA observations, but are not overly reduced to the point of being prohibitively incomplete or misleading. 
We originally explored using standard ``off the shelf'' clustering algorithms to take the $\Nsources\times\Nparams$ samples from the chain and group them into the discrete sources being fit by the model. Although not an exhaustive effort, this proved challenging due to the large dimension of parameter space, different sources located close to one another in parameter space, and the multi-modal posteriors.

A more robust approach was to group the parameters of the model by using the match between the waveforms as defined in Eq.~\ref{eq:match} and applying a match threshold $M^*$ that must be exceeded for the parameter sets to be interpreted as fitting the same source.  Seeing as it is the waveforms that are fundamentally what is being fit to the data, whereas the model parameters are just how we map from the template space to the data, clustering chain samples based on the waveform match, rather than the parameters, is naturally more effective.

The catalog production algorithm goes as follows:  
Beginning with the first sample of the chain, we compute the waveform from the parameters, produce a new \emph{Entry} to the catalog (i.e., a new discrete detection candidate), and store the chain sample in that Entry. 
The parameters and corresponding waveform become the \emph{Reference Sample} for the Entry. 
For each subsequent chain sample we again compute the waveform and check it against each catalog Entry. 
If the GW frequency of the chain sample is within 10 frequency bins of the Reference Sample we compute the match $M_{ij}$ and, if $M_{ij} > M^*$ the sample is appended to the Entry, effectively filtering all chain samples but those associated with the discrete feature in the data corresponding to the Entry. 
The check on how close the two samples are in frequency is to avoid wasteful inner-product calculations that will obviously result in $M_{ij}\sim0$. If a chain sample has been checked against all current Entries without exceeding the match threshold $M^*$ it becomes the reference sample for a new Entry in the catalog. 
Once the entire chain has been processed, the Catalog will contain many more candidate Entries than actual sources in the data (imagine a chain that has templates in the model occasionally sampling from the prior). However, the total number of chain samples in an Entry is proportional to the evidence $p(\data) = \int p(\params|\data)\, d\params$ for that candidate source. 
Thus each Entry has an associated evidence that is used to further filter insignificant features. 
The default match threshold is $M^*=0.5$ but is easily adjustable by the user.

\begin{figure}[htp]
\begin{center}
\includegraphics[width=0.5\textwidth]{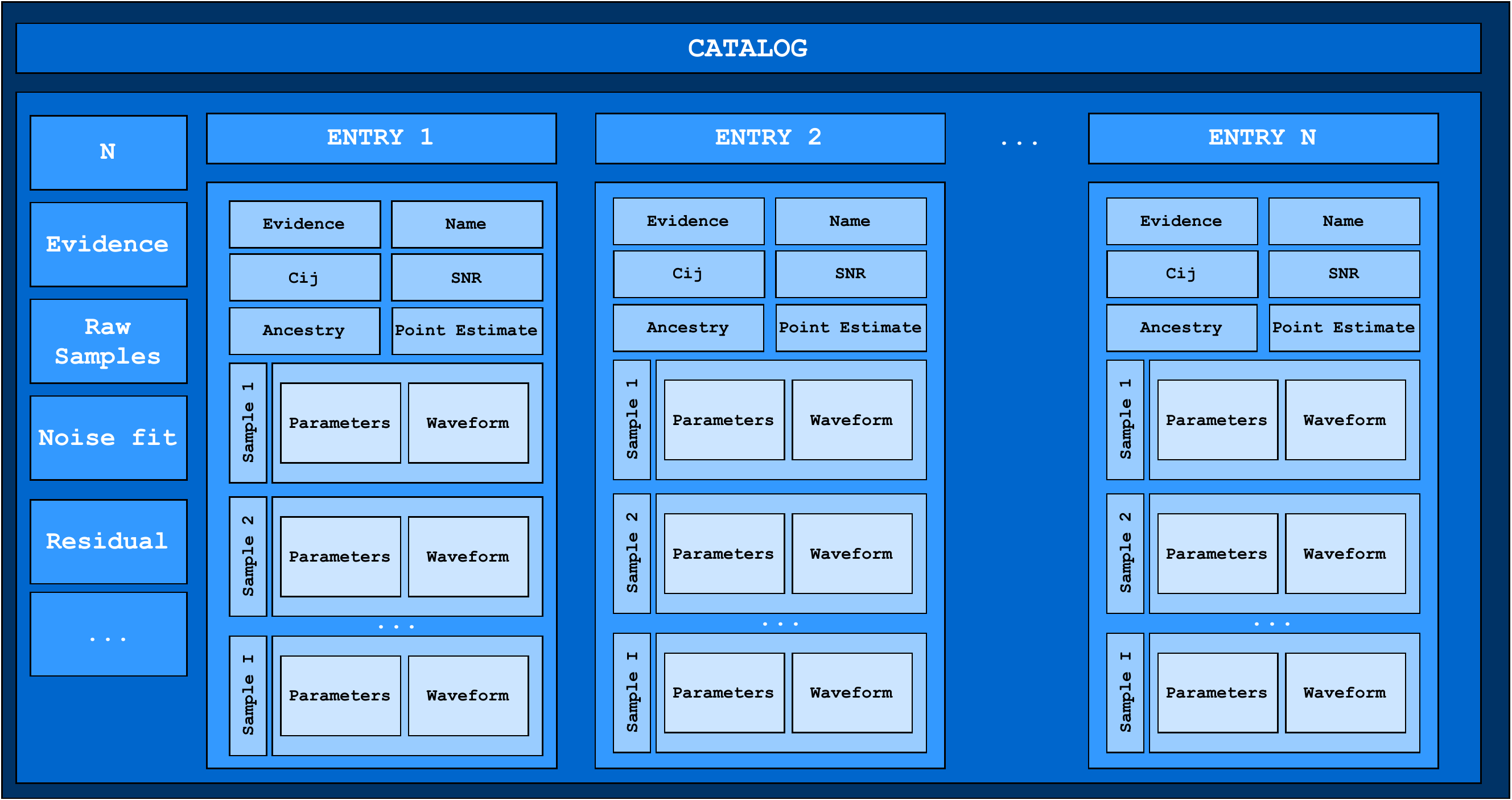}
\vspace*{-0.2in}
\caption{\label{fig:catalog} Proposed scheme for packaging chain output into higher level data products for publication in source catalogs. Raw chain output and evidences are available, as well as the posterior samples after having been filtered and clustered into discrete detected sources. Each discrete source candidate will have its own detection confidence (evidence), chain samples, point estimate, and covariance matrix error estimates so that the user can choose the most appropriate level of detail for their application of the catalog, along with metadata including the source name and history (for continuity over catalog releases), etc.}
\end{center}
\end{figure}

For each Entry, additional post-processing is then done to produce data products of varying degrees of detail depending on the needs of the end user.  We select a point-estimate as the sample containing the median of the marginalized posterior on $\f$, and store the $\SNR$, based on the reasoning that $\f$ is by far the best constrained parameter and likely the most robust way of labeling/tracking the sources.
We also compute the full multimodal $\Nparams\times\Nparams$ covariance matrix $C_{ij}$ as a condensed representation of the measurement uncertainty, and for use as a proposal when more data are acquired. From the ensemble of waveforms for each Sample in the Entry, we also compute the posterior on the reconstructed waveform. Finally, metadata about the Catalog is stored including the total number of above-threshold Entries, the full set of posterior samples, and the model evidence. 
A block diagram for the data products and how they are organized is shown in Fig.~\ref{fig:catalog}.

\subsection{Catalog Continuity}

As the observing time grows the UCB catalog will evolve.  New sources will become resolvable, marginal candidates may fade into the instrument noise, and overlapping binaries which may have been previously fit with a single template will be resolved as separate sources with similar orbital periods. 
Our scheme of identifying the binaries by their median value of $f_0$ will also evolve between releases of the catalog. 
While the association for a particular source from one catalog to the next is obvious upon inspection, the sheer number of sources requires an automated way of generating and storing the ancestry of a catalog entry in meta data.

To ensure continuity of the catalog between releases, we construct the ``family tree'' of sources in the catalog after each incremental analysis is performed. A source's ``parent'' is determined by again using the waveform match criteria, now comparing the new entry to sources in the previous catalog computed using the previous run's observing time. 
In other words, we are taking Entries found in the current step of the analysis and ``backward modeling'' the waveforms as they would have appeared during the production of the previous catalog.
The waveforms are compared to the recovered waveforms from the previous epoch to identify which sources are associated across catalogs, tracing a source's identification over the entire mission lifetime, and making it easy to quickly identify new sources at each release of the catalog.

\section{Demonstration}

To demonstrate the algorithm performance we have selected two stress-tests using data simulated for the LISA Data Challenge \emph{Radler} dataset~\footnote{\url{https://lisa-ldc.lal.in2p3.fr/ldc}}. 
The first is a high-frequency, high-$\SNR$ isolated source that challenges the convergence of the pipeline due to the many sub-dominant local maxima in the likelihood function. 
As shown in Figs.~\ref{fig:money_plot} and ~\ref{fig:logL}, new features in the algorithm have the desired affect of improving the convergence time.

\begin{figure}[htp]
\begin{center}
\includegraphics[clip=true,angle=0,width=0.5\textwidth]{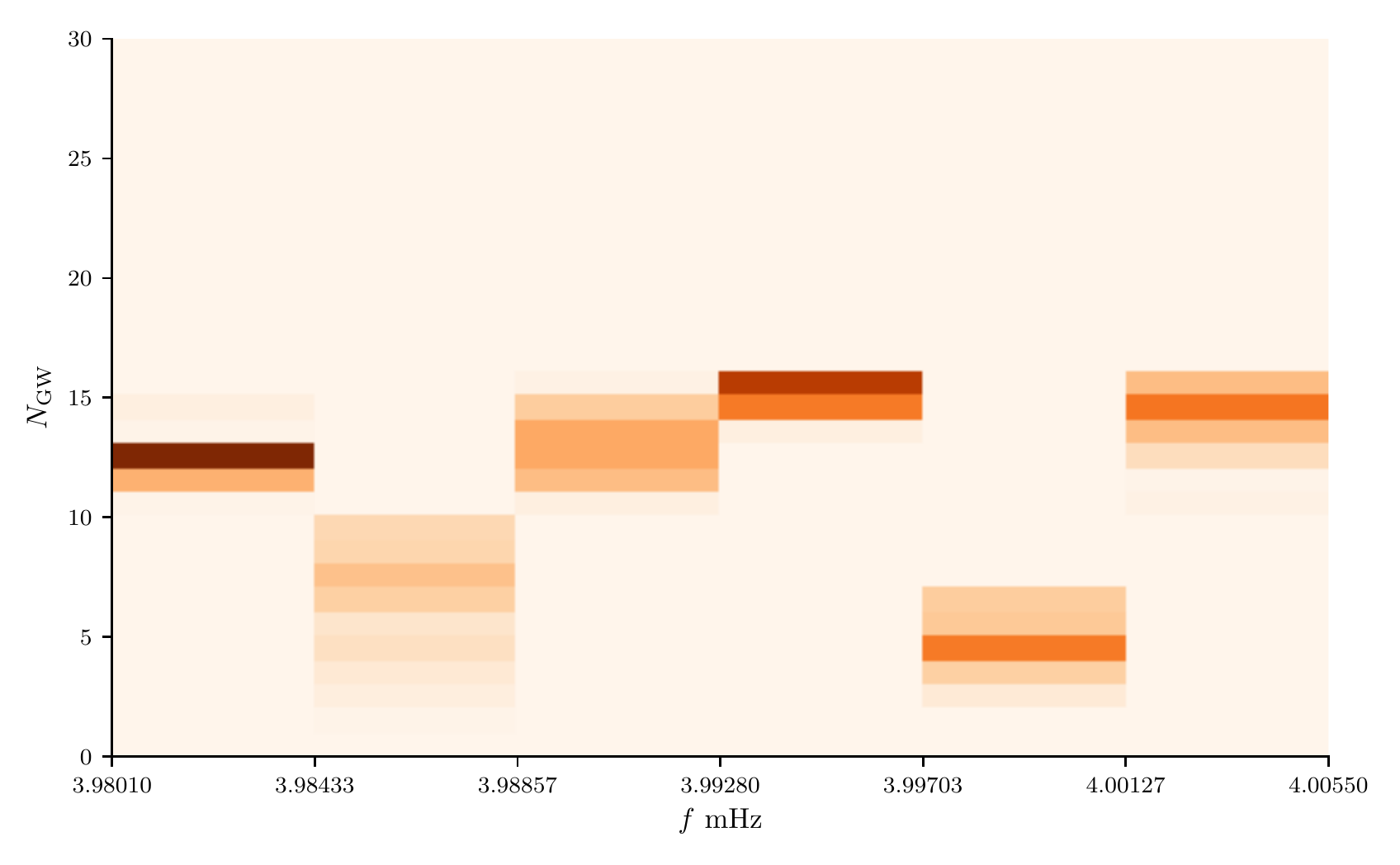} 
\end{center}
\vspace*{-0.2in}
\caption{\label{fig:4mHz-model} Heat map of posterior distribution function as a function of frequency segment and number of signals in the model.}
\end{figure}

We have also tested the pipeline on data at lower frequencies where the number of detectable sources is high, focusing on a  ${\sim}140\ \mu$Hz wide segment starting at 3.98 mHz.
The segment is subdivided into three regions to test the performance at analysis boundaries, and processed after 1.5, 3, 6, 12, and 24 months of observing. For the 24 month analysis, the full bandwidth was further divided into six regions to complete the analysis more quickly.
Fig.~\ref{fig:4mHz-model} shows a heat map of the posterior distribution function on the model dimension for the six adjacent frequency segments analyzed to cover the ${\sim}140\ \mu$Hz test segment.  The maximum likelihood model is selected for post-processing to generate a resolved source catalog.  In the event that multiple dimension models have equal likelihood the lower dimension model is selected. 

\begin{figure*}[htp]
\begin{center}
\includegraphics[clip=true,angle=0,width=1\textwidth]{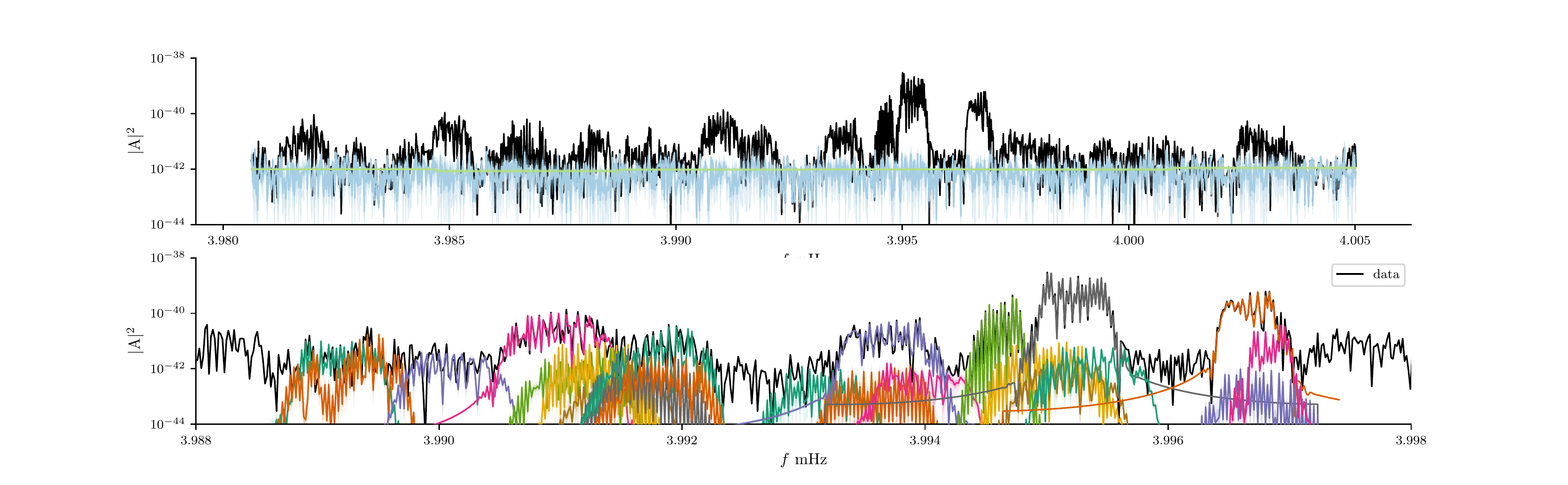}
\end{center}
\vspace*{-0.2in}
\caption{\label{fig:4mHz-waveforms} Top panel: Power spectrum for 24 months of simulated TDI-$A$ channel used to test the algorithm performance on multi-source data, with inferred residual (light blue) and noise level (green) posteriors, showing 50 and 90\% credible intervals. Bottom panel: Reconstructed waveform posteriors (using the same credible intervals) discretely identified after the 24 month analysis and post-processing zoomed in to a narrower bandwidth of the top panel, including two adjacent analysis windows.}
\end{figure*}

Fig.~\ref{fig:4mHz-waveforms} shows the data, residual, and noise model (top panel) and the posterior distributions on the reconstructed waveforms which met the criteria for inclusion into the detected source catalog after 24 months of observing (bottom panel). The waveforms, residuals, and noise reconstructions are plotted with 50\% and 90\% credible intervals, though the constraints are sufficiently tight that the widths of the intervals are small on this scale. The reconstructed waveforms are shown over a narrower-band region than the full analysis segment, containing the middle two of the six adjacent analysis windows. 

\begin{figure}[htp]
\begin{center}
\includegraphics[angle=0,width=0.5\textwidth]{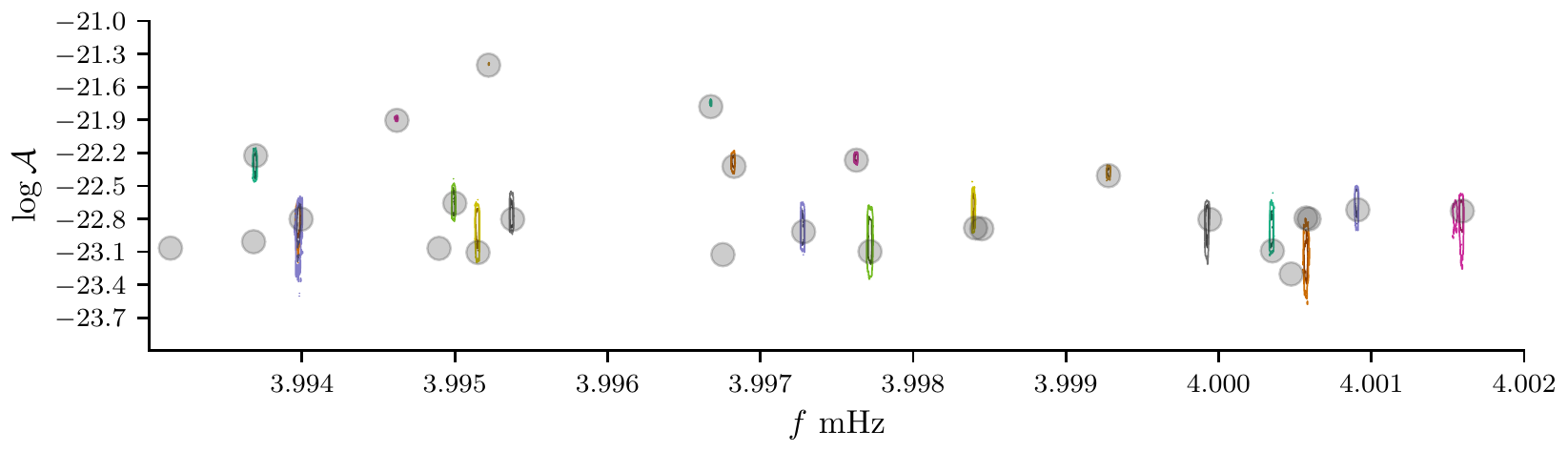}  \\
\includegraphics[angle=0,width=0.5\textwidth]{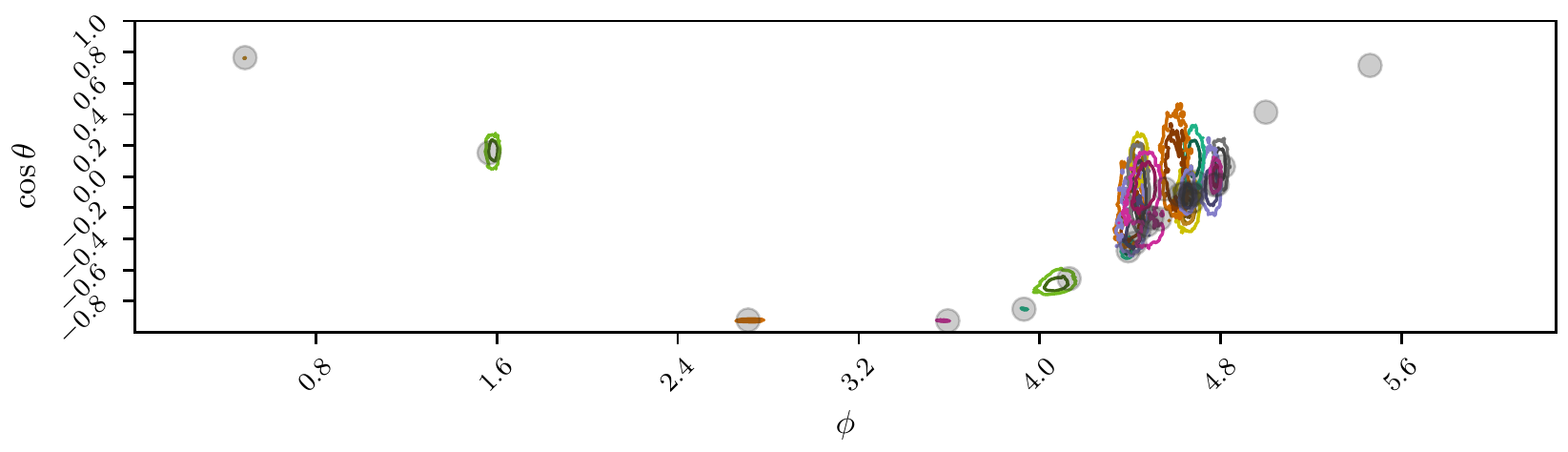}  \\
\end{center}
\vspace*{-0.2in}
\caption{\label{fig:4mHz-posteriors} Two-dimensional marginalized posteriors for a single analysis window of the full test segment of simulated data around 4 mHz after 12 months of observing time by LISA. The analysis was built up from 1.5, 3, and 6 month observations. Gray circles mark the parameter values of the injected sources. The top panel shows the frequency-amplitude plane, and the bottom panel shows the sky location in ecliptic coordinates. Contours enclose the 1 and $2\sigma$ posterior probability regions for each discrete source found in the catalog production, and the color scheme is consistent with Fig.~\ref{fig:4mHz-waveforms}.}
\end{figure}

The recovered source parameters are tested against the true values used in the data simulation and we find that our inferences about the data correspond to the simulated signals that we would expect to be detected. 
Fig.~\ref{fig:4mHz-posteriors} shows the 1- and 2-sigma contours of the marginalized 2D posteriors for the frequency-amplitude plane (top) and sky location (bottom) with gray circles marking the true parameter values. These results come from a single analysis window because the results from the full test region are overwhelming when all plotted together. 
 
\begin{figure*}[htp]
\begin{center}
\includegraphics[width=1.0\textwidth]{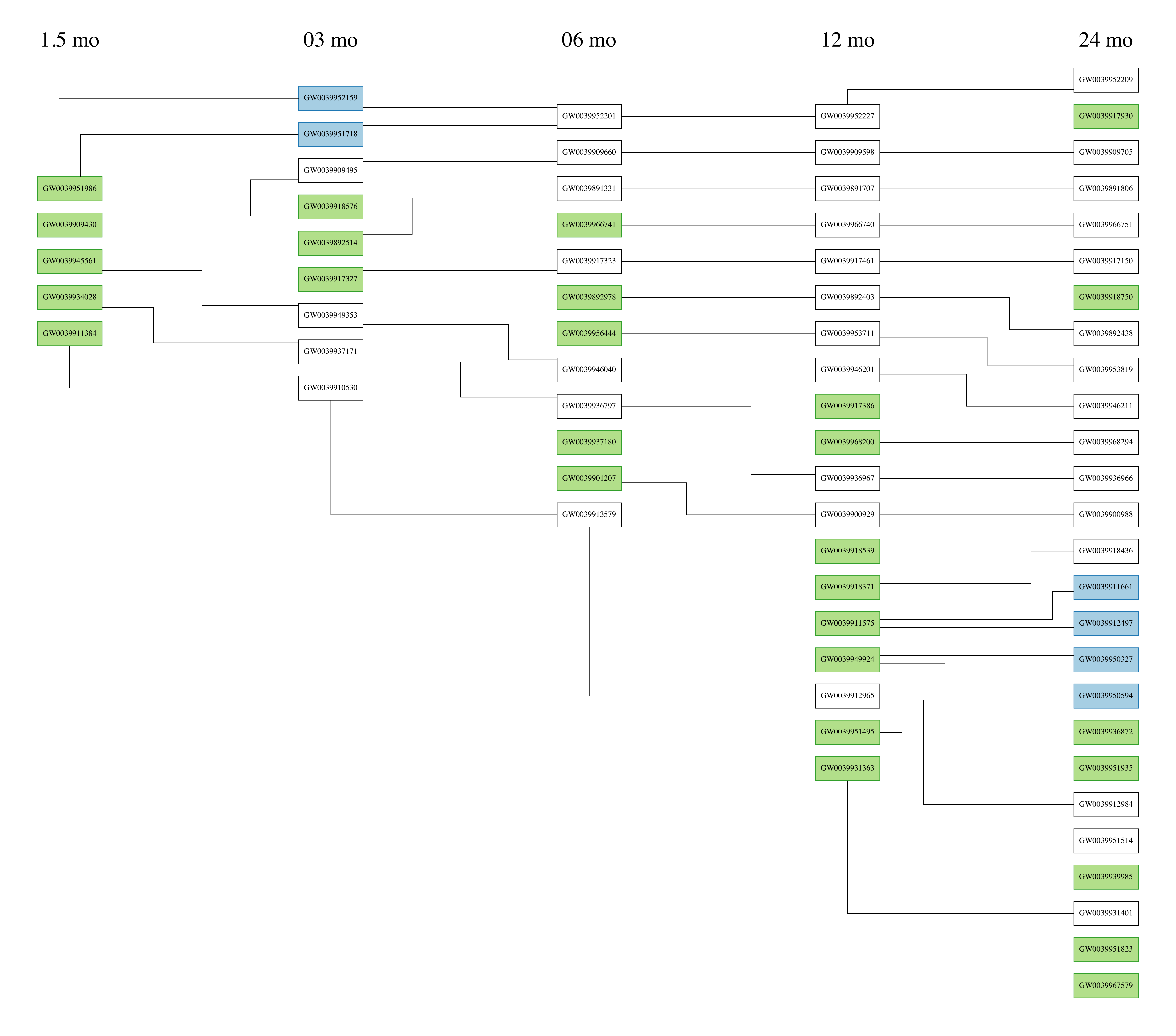}
\vspace*{-0.2in}
\caption{\label{fig:continuity} Demonstration of how catalog evolves as more data are acquired. White entries have clear ``parentage'', green are new sources in the catalog, and blue are split from a single parent.
Each entries ``geneology'' is stored as metadata in the catalog.}
\end{center}
\end{figure*}

Fig.~\ref{fig:continuity} is a graphical representation of the family tree concept for tracking how the source catalog evolves over time. 
From this diagram one can trace the genealogy of a source in the current catalog through the previous releases. The diagram is color-coded such that new sources are displayed in green, sources unambiguously associated with an entry from the previous catalog in white, and sources that share a ``parent'' with another source are in blue.

Based on the encouraging results of the narrow-band analysis shown here we will begin the analysis of the full data set. 
A thorough study of the pipeline's detection efficiency, the robustness of the parameter estimation, and optimization of MCMC and post-production settings will be presented with the culmination of the full analysis.

\section{Future Directions}

The algorithm presented here is a first step towards a fully functional prototype pipeline for LISA analysis.
We envision continuous development as the LISA mission design becomes more detailed, and as our understanding of the source population, both within and beyond the galaxy, matures.

The main areas in need of further work are: (1) Combining the galactic binary analysis with analyses for other types of sources; (2) Better noise modeling, including non-stationarity on long and short timescales; (3) Handling gaps in the data; (4) More realistic instrument response modeling and TDI generation; (5) Further improvements to the convergence time of the pipeline.

\begin{figure}[htp]
\begin{center}

\includegraphics[width=0.5\textwidth]{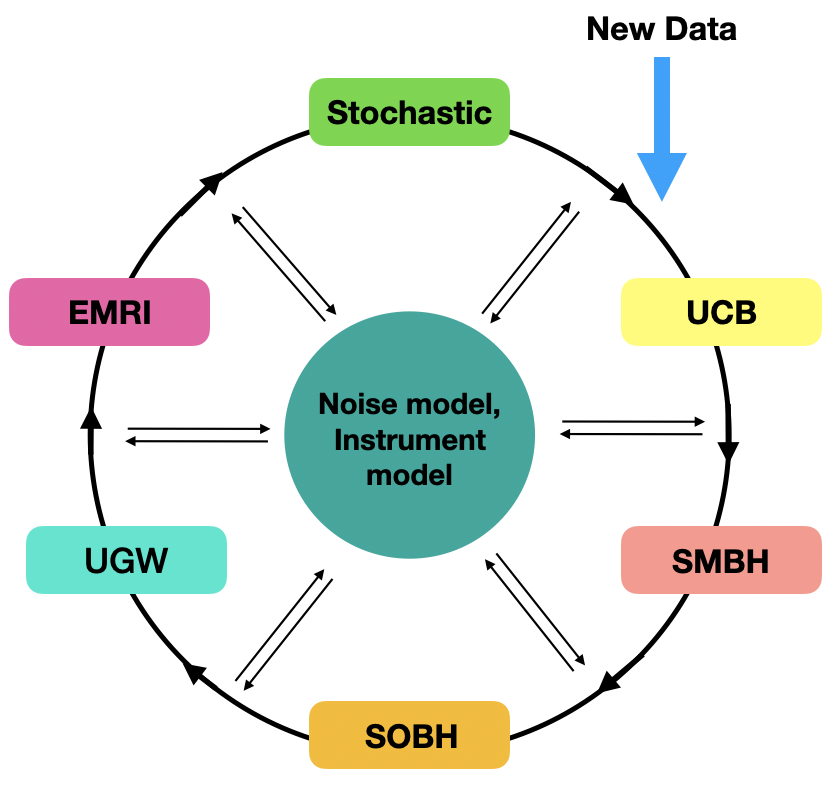}
\vspace*{-0.2in}
\caption{\label{fig:global} The UCB search as one component of a global fit. The residuals from each source analysis block are passed along to the next analysis in a sequence of Gibbs updates. New data is incorporated into the fit during the mission. The noise model and instrument models are updated on a regular basis.}
\end{center}
\end{figure}

Figure \ref{fig:global} shows one possible approach for incorporating the galactic analysis as part of a global fit. In this scheme, the analyses for each source type, such as super massive black hole binaries (SMBH), stellar origin (LIGO-Virgo) binaries (SOBH),  un-modeled gravitational waves (UGW), extreme mass ratio inspirals (EMRI), and stochastic signals (Stochastic) are cycled through, which each analysis block passing updated residuals (i.e., the data minus the current global fit) along to the next analysis block. New data is added to the analysis as it arrives.
The noise model and the instrument model (e.g., spacecraft orbital parameters, calibration parameters, etc.) are regularly updated. This blocked Gibbs scheme has the advantage of allowing compartmentalized development, and should be fairly efficient given that the overlap between different signal types is small.

A more revolutionary change to the algorithm is on the near horizon, where we will switch to computing the waveforms and the likelihood using a discrete time-frequency wavelet representation. A fast wavelet domain waveform and likelihood have already been developed~\cite{Cornish:2020}.
This change of basis allows us to properly model the non-stationary noise from the unresolved signals which are modulated by the LISA orbital motion, as well as any long-term non-stationarity in the instrument noise. 
Rectangular grids in the time-frequency plane are possible using wavelet wave packets~\cite{Klimenko_2008} which make it easy to add new data as observations continue, instead of needing the new data samples to fit into a particular choice for the wavelet time-frequency scaling, e.g. being $2^n$ or a product of primes. Wavelets are also ideal for handling gaps in the data as they have built-in windowing that suppresses spectral leakage with minimal loss of information. The time-frequency likelihood~\cite{Cornish:2020} also enable smooth modeling of the dynamic noise power spectrum $S(f , t)$ using \BayesLine type methods extended to two dimensions.

Convergence of the sampler will be improved by including directed jumps in the extrinsic parameters when using the $\mathcal{F}$ statistic proposal (as opposed to the uniform draws that are currently used). The effectiveness of the posterior-based proposals can be improved by including inter-source correlations in the proposal distributions. 
This would be prohibitively expensive if applied to all parameters as the full correlation matrix is $D \times D$, where $D=\Nsources\times\Nparams \sim 10^4$. 
However, if the sources are ordered by frequency, the $D \times D$ correlation matrix of source parameters will be band diagonal. 
We can therefore focus only on parameters that are significantly correlated, and only between sources that are close together in parameter space, while explicitly setting to zero most of the off-diagonal elements of the full correlation matrix. 
There may also be some correlations with the noise model parameters, but we do not expect these to be significant.

Along a similar vein, we will include correlations between sources in the Fisher matrix proposals. This will only be necessary for sources with high overlaps~\cite{Crowder:2004ca} which will be identified adaptively within the sampler. 
Then the Fisher matrix is computed using the parameters set $\params = \{\params_1, \params_2\}$ and waveform model ${\bf h}(\params) = {\bf h}_1(\params_1) + {\bf h}_2(\params_2)$. 

There is a large parameter space of analysis settings to explore when optimizing the computational cost of the full analysis, as well as the ``wall'' time for processing new data.  The first round of tuning the deployment strategy for the pipeline will come from studying the optimal segmenting of the full measurement band, and the cadence for reprocessing the data as the observing time increases.  

We will extend the waveform model to allow for more complicated signals including eccentric white dwarf binaries, hierarchical systems and stellar mass binary black holes which are the progenitors of the merging systems observed by ground-based interferometers~\cite{Sesana_2016}, and develop infrastructure to jointly analyze multimessenger sources simultaneously observable by both LISA and EM observatories~\cite{Korol:2017qcx,Kupfer_2018, Burdge_2019, Littenberg_2019b}.

\section*{Acknowledgments}
We thank Q. Baghi, J. Baker, C. Cutler, J. Slutsky, and J. I. Thorpe for insightful discussions during the development of this pipeline, particularly related to the catalog development. We also thank the LISA Consortium's LDC working group for curating and supporting the simulated data used in this study. TL acknowledges the support of NASA grant NNH15ZDA001N-APRA and the NASA LISA Study Office.
TR and NJC appreciate the support of the NASA LISA Preparatory Science grant 80NSSC19K0320.
KL's research was supported by an appointment to the NASA Postdoctoral Program at the NASA Marshall Space Flight Center, administered by Universities Space Research Association under contract with NASA. NJC expresses his appreciation to Nelson Christensen, Direct of Artemis at the Observatoire de la C\^{o}te d'Azur, for being a wonderful sabattical host.

\bibliography{refs}

\end{document}